\documentclass[a4paper, final, 12pt]{article}

\usepackage[utf8]{inputenc} 
\usepackage[english]{babel} 
\usepackage{eurosym} 
\usepackage{amsmath, amssymb, amsfonts, latexsym, amscd, amsthm} 
\usepackage[mathscr]{eucal} 
\usepackage{graphicx} 
\usepackage{subfig} 
\usepackage{float} 
\usepackage{color} 
\usepackage{hyperref} 

\newtheorem{proposition}{Proposition}

\theoremstyle{remark}

\usepackage{rotating} 
\usepackage{booktabs} 
\usepackage{caption} 

\usepackage{tikz} 
\usetikzlibrary{patterns} 
\usepackage{pgfplots} 
\pgfplotsset{compat=newest} 
\usepgfplotslibrary{groupplots} 
\usepackage{pgfplotstable} 

\newtheorem{theorem}{Theorem}[section]

\newtheorem{definition}{Definition}[section]

\numberwithin{equation}{section}
\newcommand{\x}{{\bf x}}
\newcommand{\y}{{\bf y}}

\begin{document}

\begin{titlepage}
   \begin{center}
       \vspace*{1cm}

        \Large   
        \begin{center}
            \textbf{Stochastic Calculus for Option Pricing}
        \end{center}
        
        \large  
        \vspace{0.5cm}
        with Convex Duality, Logistic Model, and Numerical Examination
        
       \vspace{1cm}
       \text{by}
        
       \vspace{1cm}      
       \textbf{Zheng Cao}
       
       \vspace{0.5cm}
       \text{Supervised by: Zhen-Qing Chen}

       \vfill
            
       A honors senior thesis presented for the degree of\\
       Bachelor of Science in Mathematics with Distinction
            
       \vspace{0.8cm}
     
       \includegraphics[width=0.3\textwidth]{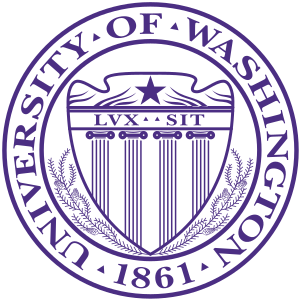}
                    
       \vspace{0.8cm}            
       Department of Mathematics \\
       University of Washington, Seattle \\
       The United States of America \\
       June 2023 
            
   \end{center}
\end{titlepage}

\newpage

\begin{abstract}

This thesis explores the historical progression and theoretical constructs of financial mathematics, with an in-depth exploration of Stochastic Calculus as showcased in the Binomial Asset Pricing Model and the Continuous-Time Models. A comprehensive survey of stochastic calculus principles applied to option pricing is offered, highlighting insights from Peter Carr and Lorenzo Torricelli's ``Convex Duality in Continuous Option Pricing Models". This manuscript adopts techniques such as Monte-Carlo Simulation and machine learning algorithms to examine the propositions of Carr and Torricelli, drawing comparisons between the Logistic and Bachelier models. Additionally, it suggests directions for potential future research on option pricing methods.

\end{abstract}

\vspace{0.5em}

\textbf{\text{Keywords:}}
Stochastic Calculus for Finance, Option Pricing,  Bachelier Model, \\ Black-Scholes-Merton Model, Logistic Model, Convex Duality, Monte-Carlo Simulation

\newpage

\tableofcontents

\newpage

\section{Introduction}

\hspace{1em}


This manuscript explores stochastic calculus for finance, with primary sources derived from Steven E. Shreve's books: ``Stochastic Calculus for Finance I: The Binomial Asset Pricing Model" \cite{SCFI} and ``Stochastic Calculus for Finance II: Continuous-Time Models" \cite{SCFII}. This undertaking also integrates valuable insights from extensive research in stock option pricing, utilizing concepts from Convex Duality, Monte Carlo Simulation, and machine learning.

The thesis delineates a concise history of the crucial inspirations and advancements in stochastic calculus for finance. This progression begins with discrete random walk models and continuous-time models. The main focus primarily examines the topics from Peter Carr and Lorenzo Torricelli's paper, ``Convex Duality in Continuous Pricing Models"\cite{ConvexDuality}. It summarizes the background, assumptions, and novel contributions of their work. It also presents fundamental probability concepts from the literature, such as martingales,  Brownian motion, and  change of measure, along with their corresponding financial applications.

A multitude of models, encompassing Bachelier, Black-Scholes-Merton, Logistic, and more, are discussed in relation to their distinct conditions and resultant outcomes. While full proofs and derivations are beyond the scope of this thesis, some references are supplied to facilitate further inquiry by interested readers.

This paper aims to equip readers with a holistic comprehension of option pricing, especially in the context of continuous-time models. Monte Carlo Simulation is employed as a practical tool for comparison and evaluation. We analyze the proposed models and suggest research potential of applying machine learning models to improve option pricing for the future. 

This thesis is submitted in partial fulfillment of the Departmental Honors requirements for Zheng Cao's undergraduate degree in Mathematics at the University of Washington. This research piece signifies the culmination of a year-long specialized program, under the mentorship and supervision of Professor Zhen-Qing Chen from the Department of Mathematics at the University of Washington.

\newpage
\section{A Brief History of Financial Mathematics} \label{sec: history}

Financial mathematics, also known as mathematical finance or quantitative finance, integrates mathematical and statistical methodologies to analyze and model financial markets and instruments. It seeks to understand and quantify the complex dynamics of economic systems, contributes to risk management, and facilitates informed investment decision-making.

The beginnings of financial mathematics are rooted in the early 20th century, gaining significant traction during the latter half of the century due to advances in computational technology and the rising complexity of financial instruments and markets. One of its key applications, which is the focus of this thesis, is option pricing.

\begin{definition}[\textbf{Option}]
    In finance, an option is a derivative financial contract that provides the buyer the right, but not the obligation, to either buy or sell an asset (such as a stock, bond, commodity, or currency) at a predetermined price (the \textit{strike price}) within a specified time frame (for an American option) or on a specified date (for a European option) \cite{hull}. The buyer pays a fee, called the \textit{premium}, to the seller for this right. The seller, or option writer, has the obligation to fulfill the contract if the buyer decides to exercise the option. 
    
    Options fall into two main categories:
    \begin{itemize}
        \item \textbf{Call Option:} Grants the holder the right to buy the underlying asset.
        \item \textbf{Put Option:} Allows the holder the right to sell the underlying asset.
    \end{itemize}
\end{definition}

Stochastic calculus, initiated by the Japanese mathematician Kiyosi Itô during World War II, is an essential tool in financial mathematics. This mathematical discipline provides a framework that accurately models and analyses the inherent randomness and unpredictability found in financial markets. It lays the foundation for the creation of advanced models that account for elements such as price fluctuations, volatility, and the impacts of unpredictable events on financial instruments. The reach of stochastic calculus extends far beyond options pricing, finding use in a wide range of finance fields, including risk management, portfolio optimization, derivative pricing, credit risk modeling, and algorithmic trading.

The European option and the American option are the two standard types of options.

\begin{definition}[\textbf{European Option}]
A European option is a financial derivative contract that can only be exercised at the time of its expiration. Let's denote $S$ as the price of the underlying asset, $K$ as the strike price, $T$ as the maturity date, and $C_E(S, T)$ and $P_E(S, T)$ as the prices of a European call and put option respectively. Then:

\begin{align*}
    C_E(S, T) &= \max(0, S - K)\\
    P_E(S, T) &= \max(0, K - S)
\end{align*}
\end{definition}

For instance, if a European call option on a stock has a strike price of \$100 and the stock price at expiration is \$110, the holder of the option can buy the stock for \$100 and sell it in the market for \$110, earning \$10 in profit.

\begin{definition}[\textbf{American Option}]
An American option is a financial derivative contract that can be exercised at any time up to its expiration. If $C_A(S, t)$ and $P_A(S, t)$ denote the prices of an American call and put option at time $t < T$ respectively, then:

\begin{align*}
    C_A(S, t) &= \max(C_E(S, T), C_E(S, t))\\
    P_A(S, t) &= \max(P_E(S, T), P_E(S, t))
\end{align*}

\end{definition}

In practice, American options offer more flexibility as they can be exercised at any time up to their expiration. This is advantageous if the holder anticipates that the price of the underlying asset will decrease before expiration. Because of this added flexibility, American options can be pricier than their European counterparts. However, particularly for call options on non-dividend-paying assets, early exercise generally presents no advantage, thus European and American options may share the same value. 

Notice, Peter Carr did not specify the option type in ``Convex Duality in Continuous Option Pricing Models", thus this thesis applies to option pricing in general; however, the data sets for the numerical examination part in Section 5 focus on American Option. 

In reality, the pricing of these options utilizes more complex models like the Black-Scholes-Merton model for European options and the binomial model or finite difference methods for American options. The precise details depend on various factors, including whether the underlying asset pays dividends, interest rates, and the volatility of the underlying asset.

In addition to European and American options, there are many other types of options, such as Asian, Barrier, Bermudan, Binary, Lookback, Rainbow, and Exotic Options. Further details on these options can be found in John C. Hull's ``Options, Futures, and Other Derivatives" \cite{hull}, and Steven E. Shreve's ``Stochastic Calculus for Finance II: Continuous-Time Models"\cite{SCFII}.

In the 1950s, the introduction of modern portfolio theory by Harry Markowitz laid the groundwork for integrating mathematical principles into finance, providing a framework for optimizing investment portfolios by balancing risk and return. This pivotal development was followed in the 1960s and 1970s by the emergence of the Black-Scholes-Merton model, which revolutionized options pricing. Leveraging the power of stochastic calculus, this model, which will be discussed further in the next section, presented a formula for pricing options.

Initially proposed by Fischer Black and Myron Scholes, the model's potential was further realized when Robert C. Merton extended their work. Merton not only provided a more rigorous mathematical derivation of the Black-Scholes formula, based on the principles of dynamic hedging and absence of arbitrage opportunities but also broadened the model's applicability by factoring in dividend payments. Beyond this, he elucidated how the formula could be harnessed within a corporate finance context. These significant contributions led to Merton sharing the 1997 Nobel Prize in Economic Sciences with Myron Scholes. Details of this model are further discussed in Section~\ref{sec: BSmodel}.

As we move to the next section of this thesis, we will explore the timeline of stochastic calculus in finance. Starting with fundamental discrete random walk models, we'll progress into the complex realm of continuous option pricing models. We'll also delve into revolutionary theories introduced by prominent scholars, including the recent work of Peter Carr. This exploration aims to provide a comprehensive understanding of the field's evolution and current state.

\newpage
\section{Stochastic Calculus for Finance} \label{sec: sto cal}

In the forthcoming section, we endeavor to traverse the established terrain of Stochastic Calculus within the financial domain, commencing with the rudimentary binomial asset pricing model, and advancing towards the more complex continuous-time models. A comprehensive analysis of the Black-Scholes model, as elucidated in the pivotal work ``The Pricing of Options and Corporate Liabilities," will also be undertaken. Furthermore, we will explore some of Peter Carr's preceding seminal contributions which form the cornerstone of his recent innovative publication, ``Convex Duality in Continuous Option Pricing Models".

\subsection{The Binomial Asset Pricing Model}


To navigate the bedrock of stochastic calculus within the finance domain, initiating our exploration with the random walk or the binomial asset pricing model proves to be an effective strategy.

    \begin{definition}[\textbf{One-dimensional Random Walk}]
          A one-dimensional random walk is a sequence of random steps along the integer number line, where each step moves one unit either to the left or to the right with equal probability. Mathematically, it can be defined as a sequence of random variables $X_1, X_2, \ldots$ such that:
    
        \begin{equation}
        S_n = \sum_{i=1}^{n} X_i
        \end{equation}
        where $S_n$ denotes the position after $n$ steps, and each $X_i$ is an independent and identically distributed random variable taking values $1$ and $-1$ each with probability $\frac{1}{2}$.  
    \end{definition}

    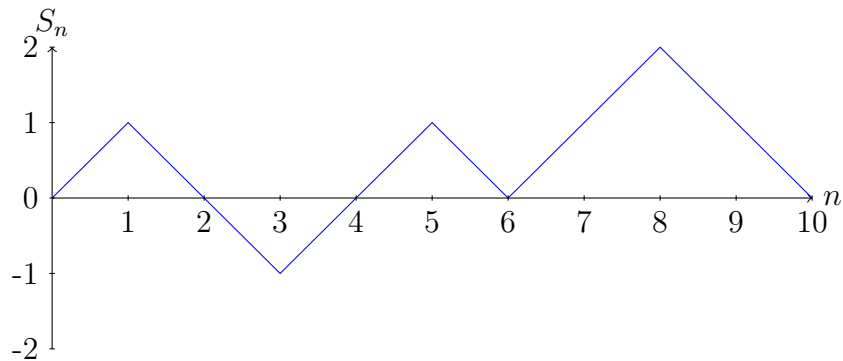
\begin{figure}[H]
        \centering
            \begin{tikzpicture}
              \draw[->] (0,0) -- (10,0) node[right] {$n$};
              \draw[->] (0,-2) -- (0,2) node[above] {$S_n$};
              
              \draw[blue] (0,0) -- (1,1) -- (2,0) -- (3,-1) -- (4,0) -- (5,1) -- (6,0) -- (7,1) -- (8,2) -- (9,1) -- (10,0);
            
              \foreach \x in {1,...,10}
                \draw (\x,1pt) -- (\x,-1pt) node[anchor=north] {\x};
              
              \foreach \y in {-2,-1,...,2}
                \draw (1pt,\y) -- (-1pt,\y) node[anchor=east] {\y};
            \end{tikzpicture}
        \caption{Example of A Random Walk}
        \label{Example of A Random Walk}
    \end{figure}

    In this diagram, the blue line represents a possible path of a random walk.
    
    \begin{theorem}
        Let $\tau_m$ be the first time a symmetric random walk reaches level m, where m is a nonzero integer. Then 
        P($\tau_m$) $<$ $\infty = 1$,
        but $\tilde{E_n}$ $= \infty$.
        
        If the RW is asymmetric then we have to deal with different cases:
        if $p_{upper}$ $>$ $1/2$, then P($\tau_m$) $<$ $\infty = 1$  and  also $\tilde{E_n} < \infty$;
        otherwise if $0 < p < 1/2$, then then P($\tau_m$) $<$ $\infty = 1$ and $\tilde{E_n} = \infty$.
        
    \end{theorem}

    \begin{definition}[\textbf{Binomial Tree Model}]
        A binomial tree model is a discrete-time (lattice-based) computational model used in finance. It is useful for deriving options prices and for illustrating how the prices of financial derivatives may evolve over time.

    \end{definition}

    This model traces its origins to the work of Cox, Ross, and Rubinstein. The binomial model works by breaking down time into a discrete number of steps, or intervals. In each interval, the price of the underlying asset can move up or down by a certain factor. This movement is represented on a tree, with each 'node' or intersection on the tree representing a possible price for the underlying asset at a given point in time.  
        
In essence, the binomial tree model is a specific type of random walk, where the random steps are proportional to the current price, rather than being a fixed amount.

Most pricing models fundamentally rest upon the assumption of a non-arbitrage environment within the financial market - a pivotal condition that is enforced in our discussion. This no-arbitrage requirement remains a key tenet in the formulation and application of these models.

    \begin{definition}[\textbf{Arbitrage}]
        An arbitrage is a portfolio value process $X(t)$ satisfying $X(0) = 0$ and also satisfying for some time $T > 0$

        \begin{equation}
            \mathbb{P}\{ X(T) \geq 0 \} = 1, \mathbb{P}\{ X(T) \geq 0 \} = 0. 
        \end{equation}
    \end{definition}

In other words, arbitrage is a trading strategy that begins with zero capital and trades in the stock and money markets in order to make money with positive probability without any possibility of losing money.

    \begin{theorem}
        The multi-period binomial model admits no arbitrage if and only if 0 $<$ d $<$ 1 + r $<$ u.
    \end{theorem}

A multi-period binomial model is introduced as a mechanic to help understand the concept. Using the example of coin tossing, we can determine the change in the stock price by the tossing outcomes. For 0 $<$ d $<$ u, we determine the price change factor to be u or d. We also consider the interest rate r of the money market, which is applied to the actions of both investing and borrowing.

    \begin{definition}[\textbf{Stock Price}]
        The stock price is the discounted risk-neutral average of its two possible prices at the next time. 
        \begin{align*}
            S_n(w_1, \ldots, w_n) &= \frac{1}{1+r} \left[ p~S_{n+1}(w_1, \ldots, w_n, H) \right. \\
            &\quad \left. + (1-p)~S_{n+1}(w_1, \ldots, w_n, T) \right]
        \end{align*}

        $S_n$ denotes the stock price at time $n$, $r$ is the risk-free rate, $H$ and $T$ are the states of the world at the next time step (indicating a move up or down, respectively), $p$ is the risk-neutral probability of the stock price moving up, and $(1-p)$ is the risk-neutral probability of the stock price moving down.
    \end{definition}

    Notice, concepts, such as the definition of the discounted risk-neutral process, will be rigorously defined in the later Continuous-Time Models chapter. 
    
    \newpage
   An example figure of a one-period binomial model is illustrated below, where $S_0$ is the initial stock price, $S_1(T) = dS_0$ represents the stock price at period 1 if the stock price decreases, and $S_1(H) = uS_0$ represents the stock price at period 1 if the stock price increases. The head ``H" and tail ``T" refer to the standard coin flip results.
    
    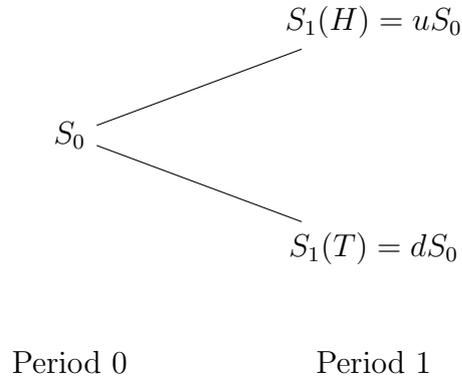
\begin{figure}[H]
        \centering
        \begin{tikzpicture}[
            grow=right,
            level distance=4cm,
            level 1/.style={sibling distance=3cm}]
        
            \node {$S_0$}
            child {node {$S_1(T) = dS_0$}}
            child {node {$S_1(H) = uS_0$}};
            
            \node at (0,-3) {Period 0};
            \node at (4,-3) {Period 1};
        \end{tikzpicture}
        \caption{One-period Binomial Model Example 1}
        \label{fig:one_period_binomial_model 1}
    \end{figure}

    As suggested by Shreve in ``Stochastic Calculus for Finance I: The Binomial Asset Pricing Model", it is common to have $d=\frac{1}{u}$. If we set $S_0 = 10, u= 1.05, and d = 0.95$, the tree below illustrates an example of a one-period binomial model\cite{SCFI}. 

        \begin{figure}[H]
        \centering
        \begin{tikzpicture}[
            grow=right,
            level distance=4cm,
            level 1/.style={sibling distance=3cm}]
        
            \node {$S_0 = 10$}
            child {node {$S_1(T) = dS_0 = 9.5$}}
            child {node {$S_1(H) = uS_0 = 10.5$}};
            
            \node at (0,-3) {Period 0};
            \node at (4,-3) {Period 1};
        \end{tikzpicture}
        \caption{One-period Binomial Model Example 1.1}
        \label{fig:one_period_binomial_model 1.1}
    \end{figure}
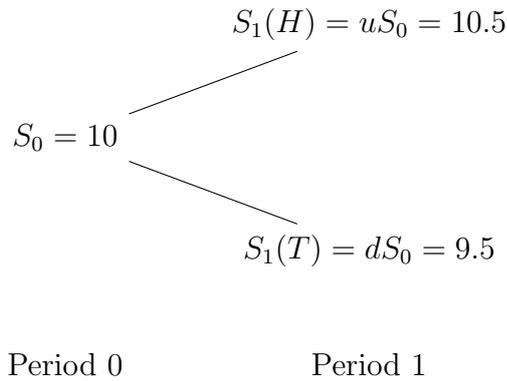

    \newpage
    A simple binomial model for multiple periods is therefore developed, for a stock price that can move up by a factor of u or down by a factor of d at each time step. For simplicity, we denote ``u" and ``d" to represent ``up" and ``down" for the stock of the given period. The initial price of the stock $S_0$ is 100, and $d= \frac{1}{u} = 0.9$:

    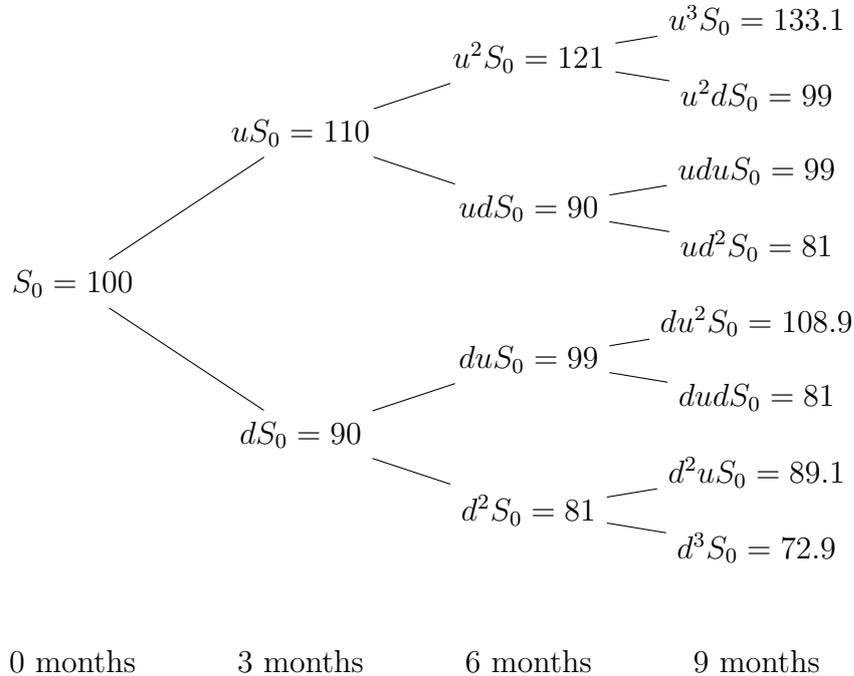
\begin{figure}[H]
        \centering
        \begin{tikzpicture}[grow=right,level distance=3cm,
                level 1/.style={sibling distance=4cm},
                level 2/.style={sibling distance=2cm},
                level 3/.style={sibling distance=1cm}]
            \node {$S_0 = 100$}
            child {node {$dS_0 = 90$}
                child {node {$d^2S_0 = 81$}
                    child {node {$d^3S_0 = 72.9$}}
                    child {node {$d^2uS_0 = 89.1$}}
                }
                child {node {$duS_0 = 99$}
                    child {node {$dudS_0 = 81$}}
                    child {node {$du^2S_0 = 108.9$}}
                }
            }
            child {node {$uS_0 = 110$}
                child {node {$udS_0 = 90$}
                    child {node {$ud^2S_0 = 81$}}
                    child {node {$uduS_0 = 99$}}
                }
                child {node {$u^2S_0 = 121$}
                    child {node {$u^2dS_0 = 99$}}
                    child {node {$u^3S_0 = 133.1$}}
                }
            };
            \node at (0,-5) {0 months};
            \node at (3,-5) {3 months};
            \node at (6,-5) {6 months};
            \node at (9,-5) {9 months};
        \end{tikzpicture}
        \caption{Example of A Binomial Tree Model}
        \label{Example of A Binomial Tree Model}
    \end{figure}

This tree is interpreting a binomial model with an up factor $u = 1.1$ and a down factor $d = 0.9$. So, at each time step, the stock price can either increase by $10\%$ or decrease by $10\%$. Each node represents the stock price at a particular time. Note that the price at a node where the stock has moved up and then down (or down and then up) is the same, since $ud = du$.

Several key contents are significant throughout the entire stochastic process studies. Among them, Martingales and Markov's Processes stand out the most.

    \begin{definition}[\textbf{Martingale}]
        A martingale is a sequence of random variables for which, at a particular time in the realized sequence, the expectation of the next value in the sequence is equal to the present observed value, regardless of all prior observed values.
        \end{definition}
    
        In mathematical terms, a discrete-time stochastic process $\{S_t\}_{t\geq0}$ is said to be a martingale if for all $t$, the conditional expectation of the next value given all past values equals the present value:
        
        \begin{equation*}
        E[S_{t+1} | S_t, S_{t-1},...,S_0] = S_t.
        \end{equation*}

In the discrete-time Binomial Asset Pricing Model, the concepts of martingales and Markov processes are fundamental. 

    \begin{definition}[\textbf{Markov Process}]
        A Markov process is a stochastic process that satisfies the Markov property, which states that the conditional probability distribution of future states of the process depends only upon the present state, not on the sequence of events that preceded it.
        \end{definition}
        
        In mathematical terms, a stochastic process $\{S_t\}_{t\geq0}$ is said to be a Markov process if for all $t$, the conditional expectation of the next value given all past values depends only on the present value:
        
        \begin{equation*}
            E[S_{t+1} | S_t, S_{t-1},...,S_0] = E[S_{t+1} | S_t].
        \end{equation*}
        
The Markov property stipulates that the value at any future time point must depend only on the current state, not on any past states. This property is inherent in the binomial model as the stock price at each node depends only on the stock price at the preceding node.

However, for the stock price process to qualify as a martingale, the expected value of the stock price at any future time point must equal its current price. The standard binomial model does not necessarily satisfy this property. To adhere to the martingale condition, the up and down factors, along with the probabilities, should be chosen such that the expected return on the stock is equal to the risk-free rate. This condition ensures the model is arbitrage-free, which is critical for it to accurately represent real-world financial markets. Arbitrage, or the opportunity to make risk-free profit without investment, is either nonexistent or rapidly exploited in real-world markets.

The Fundamental Theorem of Asset Pricing further underlines the importance of the martingale property. It states that an asset price process is a martingale if and only if the market is arbitrage-free. Thus, under the risk-neutral measure, where all assets are expected to grow at the risk-free rate of return, the asset price process in the binomial tree model is a martingale. This aligns with the principle of no-arbitrage, as an expected future value different from the current value would provide an opportunity for risk-free profit via appropriate trading.

Regarding the Martingale Representation Theorem and replicating stock option data, it's important to understand that the theorem provides a way to represent a martingale as a stochastic integral of a Brownian motion. In the context of option pricing, this could potentially allow us to replicate the payoff of an option by dynamically trading in the underlying stock and a risk-free bond, given that we can model the stock price as a stochastic process that satisfies certain conditions. This dynamic replication forms the basis for many derivative pricing models, such as the famous Black-Scholes-Merton model.

Here is the adjusted binomial tree diagram assuming $u=1/d=1.1$:

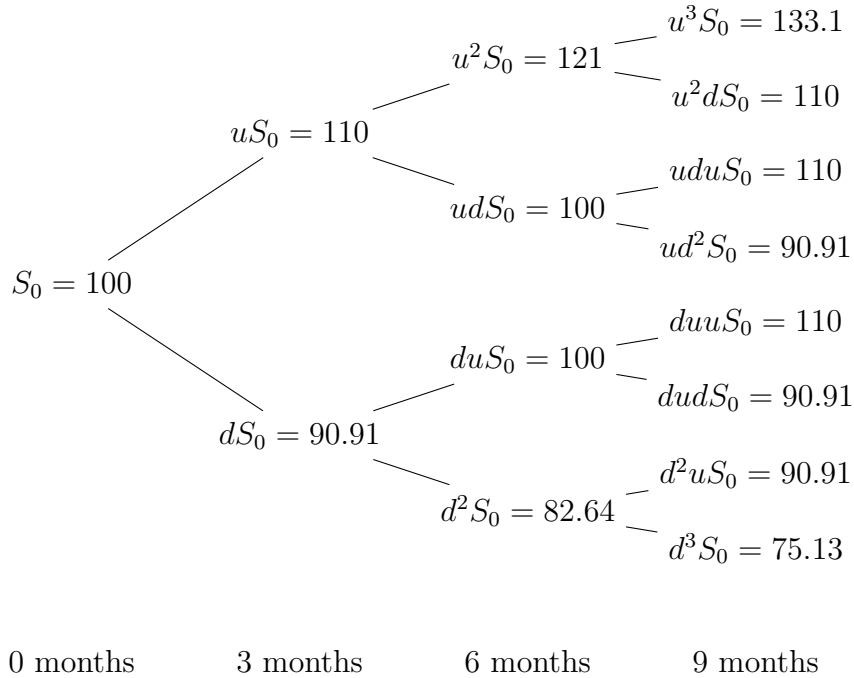
\begin{figure}[H]
    \centering
    \begin{tikzpicture}[grow=right,level distance=3cm,
            level 1/.style={sibling distance=4cm},
            level 2/.style={sibling distance=2cm},
            level 3/.style={sibling distance=1cm}]
        \node {$S_0 = 100$}
        child {node {$dS_0 = 90.91$}
            child {node {$d^2S_0 = 82.64$}
                child {node {$d^3S_0 = 75.13$}}
                child {node {$d^2uS_0 = 90.91$}}
            }
            child {node {$duS_0 = 100$}
                child {node {$dudS_0 = 90.91$}}
                child {node {$duuS_0 = 110$}}
            }
        }
        child {node {$uS_0 = 110$}
            child {node {$udS_0 = 100$}
                child {node {$ud^2S_0 = 90.91$}}
                child {node {$uduS_0 = 110$}}
            }
            child {node {$u^2S_0 = 121$}
                child {node {$u^2dS_0 = 110$}}
                child {node {$u^3S_0 = 133.1$}}
            }
        };
        \node at (0,-5) {0 months};
        \node at (3,-5) {3 months};
        \node at (6,-5) {6 months};
        \node at (9,-5) {9 months};
    \end{tikzpicture}
    \caption{Example of A Binomial Tree Model Martingale}
    \label{Example of A Binomial Tree Model Martingale}
\end{figure}

Recall, a stochastic process $X_t$ is a martingale with respect to a filtration $\mathcal{F}_t$ if the expected value of $X_t$ at any future time point given the current and past information is equal to its current value. Mathematically, this condition can be expressed as:
\begin{equation*}
    E[X_{t+1} | \mathcal{F}_t] = X_t.
\end{equation*}
Applying this to the binomial model, we note that at each node, the stock price can either move up to $uS_t$ or down to $dS_t$ with equal probability (0.5 for each). So, the expected stock price at the next time point, given the current price, equals to $S_t$,
which confirms that the process is a martingale under these conditions.

This diagram assumes that the probabilities of upward and downward moves are equal. As a result, the expected value of the stock price at each node is equal to the price at the preceding node, satisfying the martingale property. For the true risk-neutral probabilities, however, other factors such as the risk-free interest rate and the time step size may be considered, which are not included in this simplified example.

\vspace{10mm}  

Following the above assumptions and examples, we can construct the \textbf{binomial option pricing model}. It assumes that the price of the underlying asset follows a binomial tree, where it can move up or down by a certain percentage at each time step.

Consider a stock that can move up by a factor of $u$ or down by a factor of $d$ in each time step. We denote the price of the stock at time step $i$ as $S_i$. In the binomial model, the stock price evolves as follows:

\begin{equation*}
S_{i+1} = 
\begin{cases}
u \cdot S_i, & \text{with probability } p, \\
d \cdot S_i, & \text{with probability } 1-p. \\
\end{cases}
\end{equation*}

Now, consider a European call option on this stock with a strike price of $K$ and maturity of $T$. The payoff of the option at maturity is $(S_T - K)^+$, where $(x)^+$ denotes the positive part of $x$, i.e., $\max(x, 0)$.

The key idea for pricing the option is that we can form a replicating portfolio, consisting of some amount of the stock and some amount of a risk-free bond, which will have the same cash flow as the option at maturity. By the principle of no-arbitrage, the price of the option must be the same as the price of this replicating portfolio.

Once the option value is known at the final nodes, move backward through the tree. At each node, the option value is the present value of the expected future option value, assuming risk-neutral probabilities for the up and down moves. Mathematically, if $C_u$ and $C_d$ are the option values in the up and down states respectively at the next time step, $r$ is the risk-free rate, and $\Delta t$ is the time step, then the option value $C$ is given by

\begin{equation}
C = \frac{1}{1 + r \Delta t} \left( p C_u + (1 - p) C_d \right),
\end{equation}

where $p$ is the risk-neutral probability of an up move.

Continue this process until reach the initial node of the tree (i.e., today). The option value at this node is the present value of the option.

This model provides a simple yet powerful technique for replicating option pricing in discrete random walk scenarios. However, the more sophisticated Black-Scholes model extends this idea to the continuous-time setting.

\vspace{10mm}  

In summary, the Binomial Tree Model is a powerful tool for option pricing in discrete time. It is based on the assumption that the price of the underlying asset can only move up or down by a certain percentage at each time step. However, this model has its limitations. 

One significant disadvantage of the Binomial Tree Model is the high number of transactions required to dynamically hedge the option, which may be impractical in real markets due to transaction costs. As the number of time steps increases, the number of transactions also increases, leading to higher transaction costs.

Another issue arises when we consider that, as the number of steps increases and the step size decreases, the binomial tree model starts to resemble a continuous process. In fact, when the number of steps tends to infinity and the step size tends to zero, the random walk defined by the binomial tree converges to a Brownian motion. This is a key insight that motivates the transition from discrete to continuous time models for option pricing, like the Black-Scholes-Merton model.

\subsection{Continuous-Time Models}

Building upon the foundations of discrete models, the subsequent phase of our exploration within the scope of financial mathematics involves a transition to continuous-time models. The transformation from discrete random walk processes exemplified in binomial trees, to continuous stochastic processes represented by Brownian motion, introduces an added layer of complexity and mathematical rigor. ``Stochastic Calculus for Finance II: Continuous-Time Models" provides an in-depth examination of these advanced topics within financial mathematics. This includes, but is not limited to, the study of stochastic differential equations, martingale theory, stochastic integration, and the Black-Scholes-Merton model. The text presents a rigorous mathematical framework for understanding continuous-time financial models, with a notable emphasis on their application within various fields such as option pricing, risk management, and portfolio optimization. 

\subsubsection{Basics in Probability Theory}

We introduce and redefine certain terms more rigorously in the continuous-time models, starting from concepts of probability such as probability space $(\Omega, \mathcal{F}, P)$, $\sigma$ -algebra, and change of measure.

    \begin{definition}[\textbf{Probability Space}]
        A probability space is a triple $(\Omega, \mathcal{F}, P)$, where:
        \begin{itemize}
          \item $\Omega$ represents the sample space, which is the set of all possible outcomes of a random experiment.
          \item $\mathcal{F}$ denotes the event space, which is a collection of subsets of $\Omega$. It contains the events or subsets of $\Omega$ to which probabilities are assigned.
          \item $P$ is the probability measure, which is a function that assigns probabilities to events. It satisfies the following properties:
            \begin{itemize}
              \item $P(A) \geq 0$ for all $A \in \mathcal{F}$, ensuring non-negativity of probabilities.
              \item $P(\Omega) = 1$, indicating that the probability of the entire sample space is 1.
              \item For any countable sequence of disjoint events $A_1, A_2, \ldots$ (i.e., $A_i \cap A_j = \emptyset$ for $i \neq j$), we have the countable additivity property: $P\left(\bigcup_{i=1}^{\infty} A_i\right) = \sum_{i=1}^{\infty}P(A_i)$.
            \end{itemize}
        \end{itemize}

    \end{definition}
    
    Therefore, we can define both the $\sigma$-algebra and the expectation of a random variable as:
    
    \begin{definition}[\textbf{$\sigma$-algebra}]
        Let $(X, \mathcal{M}, \mu)$ be a measure space, where $X$ is a set, $\mathcal{M}$ is a $\sigma$-algebra of subsets of $X$, and $\mu$ is a measure defined on $\mathcal{M}$. For a measurable function $f: X \to \mathbb{R}$ (or $\mathbb{C}$), the Lebesgue integral of $f$ with respect to $\mu$ is denoted as $\int_X f \, d\mu$ and is defined as:
            \[
            \int_X f \, d\mu = \int_X f(x) \, d\mu(x) = \int_X f(x) \, \mu(dx),
            \]
            provided that the integral exists.
        
    \end{definition}

    \begin{definition}[\textbf{Expectation of a Random Variable}]
        Let $(\Omega, \mathcal{F}, P)$ be a probability space, and let $X: \Omega \to \mathbb{R}$ (or $\mathbb{C}$) be a random variable. The expectation or expected value of $X$ is denoted as $\mathbb{E}[X]$ and is defined as the Lebesgue integral of $X$ with respect to the probability measure $P$:
        \[
        \mathbb{E}[X] = \int_\Omega X \, dP.
        \]
    \end{definition}

    There are five vital properties for measurable functions.
    
\textbf{Properties for measurable Functions}:
    Let $f$ and $g$ be measurable functions, and let $a$ and $b$ be constants. \\
    Then, the following properties hold:
    
    \begin{enumerate}
      \item \textit{Linearity}: For any constants $a$ and $b$, we have
      \[
      \int_X (af + bg) \, d\mu = a \int_X f \, d\mu + b \int_X g \, d\mu.
      \]
      
      \item \textit{Monotonicity}: If $f(x) \leq g(x)$ for almost every $x \in X$, then
      \[
      \int_X f \, d\mu \leq \int_X g \, d\mu.
      \]
      
      \item \textit{Additivity}: If $E_1, E_2, \ldots$ are pairwise disjoint measurable sets, then
      \[
      \int_X \left(\sum_{i=1}^\infty f \chi_{E_i}\right) \, d\mu = \sum_{i=1}^\infty \int_{E_i} f \, d\mu,
      \]
      where $\chi_{E_i}$ denotes the characteristic function of the set $E_i$.
      
      \item \textit{Absolute Integrability}: If $|f|$ is integrable, then $f$ is integrable, and we have
      \[
      \left|\int_X f \, d\mu\right| \leq \int_X |f| \, d\mu.
      \]
      
      \item \textit{Change of Variables}: If $\phi: X \to Y$ is a measurable function and $f$ is integrable with respect to $\mu$ on $Y$, then
      \[
      \int_Y f \circ \phi \, d(\mu \circ \phi^{-1}) = \int_X f \, d\mu.
      \]
    \end{enumerate}

    With the settings of a probability space and a $\sigma$-algebra, we may define  $\mathcal{G}$-measurability and Borel sets. The concepts of $\mathcal{G}$-measurability and Borel sets play critical roles in measure theory and probability theory.

    \begin{definition}[\textbf{Borel Set}]
    A Borel set is any set in the smallest $\sigma$-algebra containing all open sets in a given topological space. This $\sigma$-algebra is known as the Borel $\sigma$-algebra.
    \end{definition}

    \begin{definition}[\textbf{$\mathcal{G}$-measurability}]
        Let $(\Omega, \mathcal{F}, P)$ be a probability space, and let G be a $\sigma$-algebra on $\Omega$. A random variable X defined on $(\Omega, \mathcal{F})$ is said to be $\mathcal{G}$-measurability if for every Borel set B, the pre-image $X^{-1}(B)$ belongs to G, i.e., $X^{-1}(B) \in G$.
        
        Alternatively, we can say that X is G-measurable if for all Borel sets B, the event $[X \in B]$ belongs to G, i.e., $[X \in B] \in G$.
        
        In notation, we can express the  $\mathcal{G}$-measurability of X as:
        \[
        X \text{ is G-measurable} \quad \Leftrightarrow \quad X^{-1}(B) \in G \quad \text{for all Borel sets B}.
        \]
    \end{definition}

Below is an example of Borel Sets:

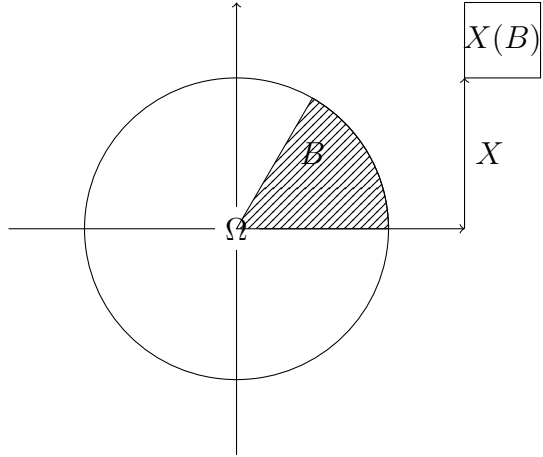
\begin{figure}[H]
    \centering
        \begin{tikzpicture}
            \draw[->] (-3,0) -- (3,0);
            \draw[->] (0,-3) -- (0,3);
            \draw (0,0) circle [radius=2cm];
            \node[fill=white] at (0,0) {$\Omega$};
            \draw[pattern=north east lines] (0,0) -- (2,0) arc [start angle=0, end angle=60, radius=2] -- cycle;
            \node at (1,1) {$B$};
            \draw[->] (3,0) -- node[right] {$X$} (3,2);
            \draw (3,2) -- (4,2) -- (4,3) -- (3,3) -- cycle;
            \node at (3.5,2.5) {$X(B)$};
        \end{tikzpicture}  
    \caption{Example of Borel Set}
    \label{Example of Borel Set}
\end{figure}

The above diagram represents a probability space $(\Omega, \mathcal{G}, P)$ and a random variable $X$ mapping a set $B$ in the $\sigma$-algebra $\mathcal{G}$ to a set $X(B)$ in the real numbers. The shaded area represents the set $B \in \mathcal{G}$ in the sample space $\Omega$. The concept of $X$ being $\mathcal{G}$-measurable is illustrated by the fact that $X$ maps the set $B$ to a set $X(B)$ on the real line.

\vspace{10mm}  

Under measure-theory, the \textbf{conditional expectation $\mathbb{E}$}, given the condition $\mathcal{G}$, of a random variable $X$ is modified as the following:

    \begin{definition}[\textbf{Conditional Expectation  $\mathbb{E}[X | \mathcal{G}]$}]
          Let $(\Omega, \mathcal{F}, P)$ be a probability space, and let $\mathcal{G}$ be a sub-$\sigma$-algebra of $\mathcal{F}$. For a random variable $X$ defined on $(\Omega, \mathcal{F})$, the conditional expectation of $X$ given $\mathcal{G}$, denoted as $\mathbb{E}[X | \mathcal{G}]$, is a random variable that satisfies the following properties:
        \begin{enumerate}
          \item $\mathbb{E}[X | \mathcal{G}]$ is $\mathcal{G}$-measurable.
          \item For any event $A \in \mathcal{G}$, $\int_A \mathbb{E}[X | \mathcal{G}] \, dP = \int_A X \, dP$.
          \item For any $B \in \mathcal{G}$, $\int_B \mathbb{E}[X | \mathcal{G}] \, dP = \int_B X \, dP$.
        \end{enumerate}      
    \end{definition}
    
    A list of additional properties of conditional expectation under measure theory is presented below, for further proof, please refer to Chapter 2 of Shreve's ``Stochastic Calculus for Finance II: Continuous-Time Models" \cite{SCFII}.
    
        \textbf{Properties of Conditional Expectation}: \label{Properties of Conditional Expectation}
        Let $X$ and $Y$ be random variables, and let $\mathcal{G}$ be a sub-$\sigma$-algebra of $\mathcal{F}$. The following properties hold for conditional expectations:
        \begin{enumerate}
          \item \textit{Linearity}: For any constants $a$ and $b$, we have
          \[
          \mathbb{E}[aX + bY | \mathcal{G}] = a\mathbb{E}[X | \mathcal{G}] + b\mathbb{E}[Y | \mathcal{G}].
          \]
          
          \item \textit{Iterated Conditioning or Tower Property}: If $\mathcal{H}$ is a sub-$\sigma$-algebra of $\mathcal{G}$, then
          \[
          \mathbb{E}\left[\mathbb{E}[X | \mathcal{G}] | \mathcal{H}\right] = \mathbb{E}[X | \mathcal{H}].
          \]
          
          \item \textit{Taking Out What is Known}: If $Y$ is $\mathcal{G}$-measurable, then
          \[
          \mathbb{E}[XY | \mathcal{G}] = Y\mathbb{E}[X | \mathcal{G}]
          \].
          
          \item \textit{Law of Total Expectation}: If $\mathcal{G}_1, \mathcal{G}_2, \ldots, \mathcal{G}_n$ form a partition of $\mathcal{F}$, then
          \[
          \mathbb{E}[X] = \sum_{i=1}^n \mathbb{E}[X | \mathcal{G}_i]P(\mathcal{G}_i).
          \]

        \item \textit{Jensen's Inequality}: If $X$ is an integrable random variable and $g: \mathbb{R} \to \mathbb{R}$ is a convex function, then
          \[
          g\left(\mathbb{E}[X | \mathcal{G}]\right) \leq \mathbb{E}[g(X) | \mathcal{G}].
          \]

        \end{enumerate}
    These properties are significant for the modeling and analysis in the following sections.

\subsubsection{Change of Measure}\label{sec: Change of Measure}

A \textbf{change of measure}, or transformation of measure, is a technique that is commonly employed in the field of measure theory. It is a process that allows us to switch between two measures. This technique is extensively applied in the realm of probability theory and financial mathematics, particularly in the computation of the prices of financial derivatives.

\begin{definition}[\textbf{Change of Measure}] \label{def: Change of Measure}

Consider a probability space $(\Omega, \mathcal{F}, \mathbb{P})$. Let $\tilde{\mathbb{P}}$ be another probability measure on $(\Omega, \mathcal{F})$, and let $Z$ be an almost surely positive random variable that relates $\tilde{\mathbb{P}}$ and $\mathbb{P}$ via:

    \begin{equation}
        \tilde{\mathbb{P}}(A) = \int_A Z(w)d\tilde{\mathbb{P}}(w)
    \end{equation}

with $\mathbb{E}Z = 1$.

Then $Z$ is called the Radon-Nikodym derivative (also known as the likelihood ratio) of $\tilde{\mathbb{P}}$ with respect to $\mathbb{P}$:

\begin{equation}
    Z = \frac{d\tilde{\mathbb{P}}}{d\mathbb{P}}.
\end{equation}

\end{definition}

\begin{theorem}[\textbf{Radon-Nikodym}] \label{Radon-Nikodym}
    Let $\mathbb{P}$ and $\tilde{\mathbb{P}}$ be equivalent probability measures defined on $(\Omega, \mathcal{F})$. Then there exists an almost surely positive random variable $Z$ such that $\mathbb{E}Z = 1$ and \\
        $\tilde{\mathbb{P}}(A) = \int_A Zd\mathbb{P}(w)$ for every $A \in \mathcal{F}$.
\end{theorem}

The concept of a ``change of measure" can be illustrated as a transformation of the measure of a certain set in the given measure space. Imagine we have a universe $\Omega$, in which there is a particular subset $A$. In the context of measure theory, this subset can be associated with different ``sizes" or ``weights", depending on the measure we apply. 

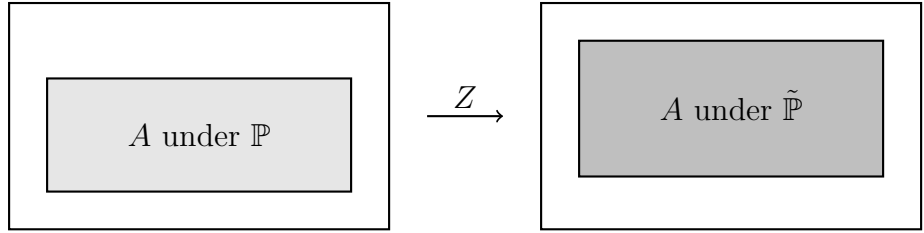
\begin{figure}[H]
    \centering
    \begin{tikzpicture}
    \draw [thick] (0,0) rectangle (5,3);
    \node at (2.5,1.5) {$\Omega$};
    \draw [thick,fill=gray!20] (0.5,0.5) rectangle (4.5,2);
    \node at (2.5,1.25) {$A$ under $\mathbb{P}$};
    \draw [->,thick] (5.5,1.5) -- (6.5,1.5);
    \node at (6,1.75) {$Z$};
    \draw [thick] (7,0) rectangle (12,3);
    \node at (9.5,1.5) {$\Omega$};
    \draw [thick,fill=gray!50] (7.5,0.7) rectangle (11.5,2.5);
    \node at (9.5,1.65) {$A$ under $\tilde{\mathbb{P}}$};
    \end{tikzpicture}
    \caption{The visual representation of a change of measure from $\mathbb{P}$ to $\tilde{\mathbb{P}}$. The set $A$ within the universe $\Omega$ has different ``sizes" under $\mathbb{P}$ and $\tilde{\mathbb{P}}$, represented by the different shades of gray.}
\end{figure}

In the above illustration, we see the same universe $\Omega$ and the same subset $A$ under two different measures, $\mathbb{P}$ and $\tilde{\mathbb{P}}$. However, the ``size" of set $A$ under these two measures is different, represented by varying shades of gray. By using a change of measure, we essentially change our perspective on the size or importance of the set $A$ within $\Omega$. This provides a powerful tool in measure theory, allowing us to navigate through different probability spaces with relative ease.

\subsubsection{Brownian Motion}

\textbf{Scaled random walks}, denoted as $W^n(t)$, are stochastic processes that can be defined as discrete-time approximations of Brownian motion. They are obtained by scaling and summing a sequence of independent and identically distributed random variables over a fixed time interval.

Let's consider a sequence of independent and identically distributed random variables $X_1, X_2, X_3, \ldots$ with mean $\mu$ and variance $\sigma^2$. The scaled random walk $W^n(t)$ is defined as:

\[
W^n(t) = \frac{1}{\sqrt{n}} \sum_{i=1}^{[nt]} X_i,
\]

where $n$ represents the number of steps in the random walk and $[nt]$ denotes the integer part of $nt$. The scaling factor $\frac{1}{\sqrt{n}}$ ensures that as $n$ increases, the random walk converges to Brownian motion.

To visualize a scaled random walk, let's consider an example where $X_i$ follows a standard normal distribution ($\mu = 0$ and $\sigma = 1$). We choose a time interval of $t = 1$ and set $n = 100$.

We generate a sequence of independent standard normal random variables $X_1$, $X_2$, $\ldots$, $X_{100}$ and compute the scaled random walk $W^{100}(t)$ using the formula mentioned earlier. Here is a plot illustrating the trajectory of the scaled random walk:

    \begin{center}
        \includegraphics[width=0.6\textwidth]{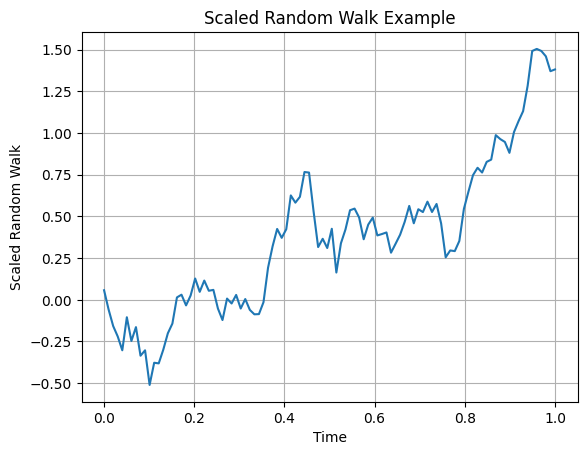}
    \end{center}

As indicated above, the scaled random walk exhibits a path that resembles the behavior of a Brownian motion. It demonstrates a random, continuous, and fluctuating trajectory over time, which is related to o Brownian Motion.

Brownian motion can be obtained as the limit of scaled random walks as the number of steps $n$ tends to infinity. As $n$ increases, the scaling factor $\frac{1}{\sqrt{n}}$ diminishes, resulting in finer increments in the random walk. In the limit, the increments become infinitesimal, leading to the continuous and unpredictable behavior of Brownian motion.

Mathematically, we have:

\[
W(t) = \lim_{n \to \infty} W^n(t),
\]

where $W(t)$ represents the Brownian motion. This connection highlights the relationship between scaled random walks and the continuous-time stochastic process of Brownian motion.

\vspace{10mm}  

As we traverse the landscape of financial mathematics, the journey naturally progresses from discrete to continuous models. The preceding sections provided an introduction to the foundations of the discrete models -- random walks and binomial trees -- as well as their applications in the financial domain. Having developed a robust understanding of these concepts, we now turn our attention to the intricacies of continuous models. To make this leap, we rely heavily on the mechanics of measure theory. This leads us to a pivotal concept within the continuous domain, the continuous-time stochastic process, also known as Brownian motion. As we will see, Brownian motion plays a central role in the development of option pricing within continuous-time models. Thus, understanding its characteristics and implications is crucial for the comprehensive study of modern financial mathematics.

\begin{definition}[\textbf{Brownian Motion}]
Let $(\Omega, \mathcal{F}, \mathcal{F}_t, P)$ be a probability space. For each $w \in \Omega$, suppose there is a continuous function $W(t), t \geq 0$ that satisfies $W(0) = 0$ and that depends on $w$. Then $W(t), t \geq 0$, is a Brownian motion (BM) if for all $0 = t_0 < t_1 < \dots < t_m$ the increments
\begin{gather*}
W(t_1) = W(t_1) - W(t_0), W(t_2) - W(t_1), \dots, W(t_m) - W(t_{m-1}) \\
\text{are independent and each of these increments is normally distributed with} \\
\mathbb{E}[W(t_{i+1}) - W(t_i)] = 0,\\
Var[W(t_{i+1}) - W(t_i)] = t_{i+1} - t_i.
\end{gather*}
\end{definition}

    The example below illustrates the concepts of Brownian Motion, which suggests some of its basic properties.

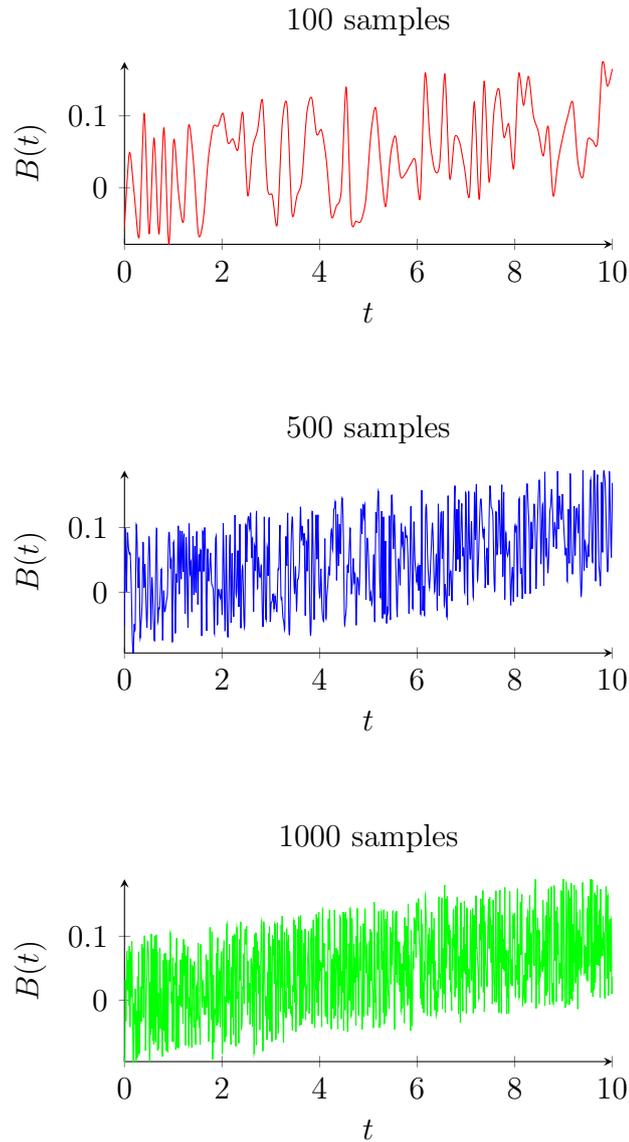
\begin{figure}[H]
    \centering
    \begin{tikzpicture}
        \begin{groupplot}[
            group style={group size=1 by 3, vertical sep=3cm},
            height=4cm, width=8cm,
            domain=0:10, 
            axis lines = left,
            xlabel = $t$,
            ylabel = {$B(t)$},
        ]
        \nextgroupplot[samples=100, title={100 samples}]
        \addplot [color=red, mark=none, smooth] {rand*0.1 + 0.01*x};
        \nextgroupplot[samples=500, title={500 samples}]
        \addplot [color=blue, mark=none, smooth] {rand*0.1 + 0.01*x};
        \nextgroupplot[samples=1000, title={1000 samples}]
        \addplot [color=green, mark=none, smooth] {rand*0.1 + 0.01*x};
        \end{groupplot}
    \end{tikzpicture}
    \caption{Approximation of Brownian motion with random sample points.}
\end{figure}

3 different approximations of Brownian motion plots are presented, each with a random walk from t=0 to t=10, with 100, 500, and 1000 sample points per plot. For each point, it generates a random number, multiplies it by 0.1 to decrease the step size, and then adds a small deterministic drift of 0.01*x. Notice, this method of visualization is quite simplistic and doesn't accurately represent the behavior of Brownian motion.

Now that we have built an understanding of the fundamental characteristics and visualization of Brownian motion, let us delve into one of its essential properties: Quadratic Variation. This property provides additional insight into the nature of Brownian motion and sets it apart from other stochastic processes. 

\begin{definition}[\textbf{Quadratic Variation}]
    The quadratic variation of a Brownian motion $(W_t)_{t \geq 0}$ over a partition $0 = t_0 < t_1 < t_2 < \ldots < t_n = T$ is defined as:
        \[
        \langle W \rangle_T = \lim_{\|\Pi\| \to 0} \sum_{i=1}^{n} (W_{t_i} - W_{t_{i-1}})^2,
        \]
        where $\|\Pi\|$ denotes the mesh size of the partition and the convergence is in probability.

\end{definition}
        
        The property of Brownian motion states that for any $T \geq 0$, the quadratic variation $\langle W \rangle_T$ is equal to $T$. In other words, the quadratic variation of Brownian motion is deterministic and grows linearly with time. It is a key characteristic that distinguishes Brownian motion from other stochastic processes.

The \textbf{reflection principle} is a fundamental result in the theory of Brownian motion that relates the probabilities of the process reaching certain levels. In the context of the reflection principle, we consider a standard Brownian motion denoted by $W(t)$, and the following statement holds:

\begin{theorem}[\textbf{Reflection Equality}]
    \[P(\tau_m \leq t, W(t) \leq w) = P(W(t) \geq 2m - w), \quad w \leq m, \ m > 0.\].
\end{theorem}
In this equation:
    \begin{itemize}
    \item $P(\tau_m \leq t, W(t) \leq w)$ represents the probability that the Brownian motion $W(t)$ hits the level $w$ or lower before time $t$, given that it hits or crosses the level $m$ at some point before time $t$. Here, $\tau_m$ denotes the first passage time of $W(t)$ at level $m$.
    \item $P(W(t) \geq 2m - w)$ represents the probability that the reflected Brownian motion, obtained by taking the absolute value of $W(t)$, exceeds or equals the level $2m - w$ at time $t$.
    \end{itemize}

The principle states that the probability of the reflected Brownian motion exceeding or equaling a certain level at time $t$ is equal to the probability of the original Brownian motion reaching the same level at time $t$, given that it has already hit or crossed the level $m$. The visualization below provides an example illustrating the reflection equality with a simulated Brownian path.

    \begin{figure}[H]
        \centering
        \includegraphics[width=0.6\textwidth]{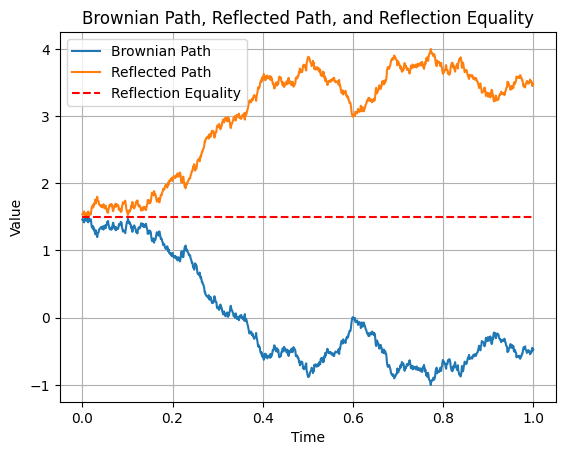}
        \label{Brownian Path, Reflected Path, and Reflection Equality}
    \end{figure}

\vspace{10mm}

Recall, for a standard Brownian Motion,  the random variable $W_t$ at any fixed time $t$ is normally distributed in general. We hereby introduce a modified Brownian Motion, the Geometric Brownian Motion.

\begin{definition}[\textbf{Geometric Brownian Motion (GBM)}]
    
    A geometric Brownian motion $S_t$ is a stochastic process given by the following stochastic differential equation:
    \begin{equation}
        \mathrm{d}S_t = \mu S_t \mathrm{d}t + \sigma S_t \mathrm{d}W_t,
    \end{equation}
    where $W_t$ is a standard Brownian motion, and $\mu$ and $\sigma$ are constants representing the drift and volatility, respectively. The solution to the equation is:
    \begin{equation}
        S_t = S_0 \exp \left( (\mu - \frac{1}{2} \sigma^2) t + \sigma W_t \right).
    \end{equation}
\end{definition}

Because of the exponential function in its definition, the random variable $S_t$ at any fixed time $t$ follows a log-normal distribution.

In contrast to a standard Brownian motion, which can take on any real value, a geometric Brownian motion can only take on positive real values due to the exponential function. Moreover, the increments of a geometric Brownian motion are not independent. Instead, it is the logarithmic returns, or continuously compounded returns, that are independent and normally distributed, a property that makes geometric Brownian motion suitable for modeling asset prices in financial mathematics. 

While a Brownian motion follows a normal distribution, a geometric Brownian motion follows a log-normal distribution, due to the multiplicative nature of its changes over time. The financial applications of GBM are further discussed in the context of the Black-Scholes model in Section~\ref{sec: BSmodel}.

\vspace{10mm}  

In this thesis, our focus transitions to a specialized mathematical framework, known as Stochastic Differential Equations (SDEs), a potent tool used to capture the dynamic nature of random variables over time. Unlike their counterparts in PDEs or Ordinary Differential Equations (ODEs), SDEs are uniquely equipped to tackle the inherent uncertainties prevalent in financial systems. Among the diverse array of stochastic processes, the Itô process holds particular significance, forming the bedrock of Stochastic Calculus. In the forthcoming sections, we delve into the intricacies of Stochastic Calculus and its consequential role in financial applications, particularly in the realm of option pricing.

    \begin{definition}[\textbf{Itô Process (Stochastic Process)}]
        An Itô process, also known as a stochastic process, is a mathematical model that describes the evolution of a random variable over time. It is defined by the following stochastic differential equation (SDE):
        \[
        dX_t = \mu(t, X_t) \, dt + \sigma(t, X_t) \, dW_t,
        \]
        where $X_t$ represents the Itô process at time $t$, $\mu(t, X_t)$ is the drift coefficient, $\sigma(t, X_t)$ is the diffusion coefficient, and $dW_t$ denotes the differential of a standard Brownian motion.
    \end{definition}

        The Itô process is characterized by the integration of the drift term $\mu(t, X_t)$ with respect to time $t$ and the diffusion term $\sigma(t, X_t)$ with respect to the Brownian motion $W_t$. It captures both deterministic and random components, allowing for modeling various phenomena subject to both systematic and random influences.
        
        The solution to the Itô process is given by the stochastic integral, known as the Itô integral. It provides a framework for studying and analyzing the behavior of stochastic processes and is widely used in mathematical finance, physics, and other fields dealing with stochastic phenomena.

A \textbf{diffusion process} is a stochastic process that describes the random movement of a quantity over time, where the increments of the process are normally distributed. It is characterized by continuous and smooth trajectories, exhibiting a continuous-time analog of Brownian motion. One commonly used diffusion process is the geometric Brownian motion. which is a key component of the Black-Scholes model for option pricing.

 A \textbf{Brownian Bridge} is a stochastic process that represents a Brownian motion over a specified interval while fixing the endpoints. It is constructed such that the process starts at a given value at one endpoint and ends at a different value at the other endpoint, following a continuous and random path in between.

    We can compare the paths of three stochastic processes: Brownian motion, geometric Brownian motion, and Brownian bridge. The trajectory of \textbf{Brownian motion}, a random walk with normally distributed increments, shows a more erratic and unpredictable movement, with the increments at each time step represented by the changes in the vertical direction. The trajectory of \textbf{geometric Brownian motion}, a continuous-time stochastic process commonly used to model the behavior of stock prices, incorporates a drift term and a volatility term, resulting in a smoother and upward-biased movement due to the exponential growth factor. The trajectory of \textbf{Brownian bridge}, which is a modification of Brownian motion, starts and ends at the same value and is conditioned to pass through a specific point (in this case, the midpoint of the time interval). This constraint leads to a more controlled and less erratic movement compared to Brownian motion.
    
     Brownian motion captures the random and unpredictable nature of price movements, while geometric Brownian motion provides a framework for modeling exponential growth with stochastic fluctuations. Brownian bridge introduces additional constraints to create smoother paths, making it useful in situations where a specific boundary condition needs to be satisfied.

\subsubsection{Stochastic Calculus}

In this section, we aim to provide a comparison of the Riemann, Lebesgue, and Itô integrals. Each of these integrals has distinct assumptions, and notations, and is used in various fields of mathematical analysis. We also emphasize both the Partial Differential Equation (PDE) and Stochastic Differential Equation (SDE) in the context of finance.

First, what is Itô Integral?

    \begin{definition}[\textbf{Itô Integral}]
        
        The Itô Integral is a stochastic integral used to define the integral of a stochastic process with respect to a stochastic process. Let $W_t$ be a standard Brownian motion and $f(t, \omega)$ be an adapted process satisfying suitable conditions. The Itô Integral of $f(t, \omega)$ with respect to $W_t$ is denoted as $\int_{0}^{T} f(t, \omega) dW_t$ and is defined as the limit of the following sequence:
        \[
        \int_{0}^{T} f(t, \omega) dW_t = \lim_{|\Pi| \to 0} \sum_{i=0}^{n-1} f(t_i, \omega)[W(t_{i+1}) - W(t_i)],
        \]
        where $|\Pi|$  denotes the mesh size of the partition $\Pi$ and  $n$ is the number of sub-intervals and the convergence is in probability.
    \end{definition}    
    
        The Itô Integral satisfies the following properties:
    \newline       
    \textbf{Properties of Itô Integral}
        \begin{itemize}
          \item \textbf{Linearity}: For any constants $a$ and $b$, and adapted processes $f(t, \omega)$ and $g(t, \omega)$, we have
          \[
          \int_{0}^{T} [af(t, \omega) + bg(t, \omega)] dW_t = a\int_{0}^{T} f(t, \omega) dW_t + b\int_{0}^{T} g(t, \omega) dW_t
          \]
          
          \item \textbf{Isometry}: For any adapted process $f(t, \omega)$, the following holds:
          \[
          \mathbb{E}\left[\left(\int_{0}^{T} f(t, \omega) dW_t\right)^2\right] = \mathbb{E}\left[\int_{0}^{T} f(t, \omega)^2 dt\right].
          \]
        \end{itemize}

        \textbf{Example}: Consider the Itô process $X_t = \int_{0}^{t} f(s) dW_s$, where $f(s)$ is an adapted process. We can compute the Itô Integral explicitly for a simple case. Let $f(s) = s$, and $T > 0$ be a fixed time. The Itô Integral of $f(s)$ with respect to $W_s$ is given by:
        \[
        \int_{0}^{T} f(s) dW_s = \int_{0}^{T} s dW_s = \frac{1}{2}W_T^2 - \frac{1}{2}T,
        \]
        where $W_T$ denotes the value of the Brownian motion $W_t$ at time $T$.
        
\vspace{10mm}  

\newpage
Consider the concept of approximating a continuously varying integrand using the Ito integral in stochastic mathematics for finance. In this example, we generate a time axis and simulate a continuously varying integrand with \textbf{jumps}.

The \textbf{integrand} is defined as the sum of a sinusoidal function with added random noise. We accumulate the values of the integrand to simulate jumps, and then multiply by the square root of the time increment to obtain the approximation of the Ito integral.

    \begin{figure}[H]
        \centering
        \includegraphics[scale=0.9]{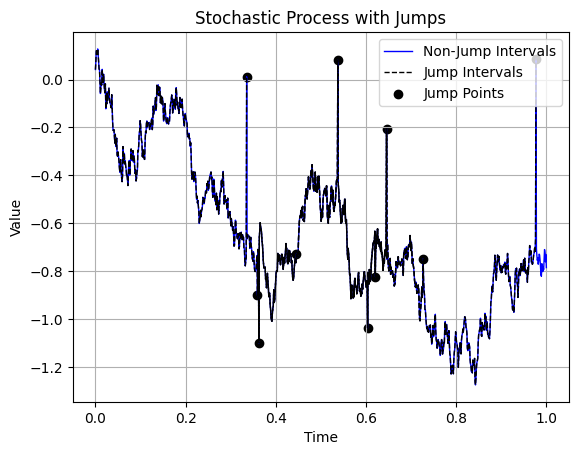}
        \caption{Stochastic Process with Jumps}
        \label{Stochastic Process with Jumps}
    \end{figure}
    
The resulting plot displays the continuously varying integrand and its approximation over time. It shows the fluctuating behavior of the integrand with jumps and the approximation of the Ito integral.

\vspace{10mm}  

We hereby compare and contrast three common integrals: the Riemann integral, the Lebesgue integral, and the Itô integral.

The \textbf{Riemann integral} is defined for a function of a real variable on a closed interval $[a, b]$. The integral of a function $f$ from $a$ to $b$ is denoted as $\int_a^b f(x) dx$, where $dx$ is an infinitesimally small increment in the variable $x$. 

The \textbf{Lebesgue integral} extends the idea of the Riemann integral to include more general classes of functions and measures. It is defined on a measure space and works particularly well for functions that are not well-behaved at a countable number of points.

The \textbf{Itô integral} is a modification of the Riemann and Lebesgue integrals in the context of stochastic calculus. The Itô integral of a stochastic function $f(t, B(t))$ with respect to a Brownian motion $B(t)$ is denoted as $\int_0^T f(t, B(t)) dB(t)$.

\renewcommand{\arraystretch}{1.5} 
\begin{table}[H]
\centering
\begin{tabular}{|l|l|l|l|}
\hline
 & Riemann & Lebesgue & Itô \\
\hline
Definition & Defined on a & Defined on a & Defined on a \\
 & closed interval & measure space & stochastic process \\
\hline
Notation & $\int_a^b f(x) dx$ & $\int_E f d\mu$ & $\int_0^T f(t, B(t)) dB(t)$ \\
\hline
Usage & Standard calculus & Measure theory & Stochastic calculus \\
\hline
\end{tabular}
\caption{Comparison of Riemann, Lebesgue, and Itô Integrals}
\end{table}
\renewcommand{\arraystretch}{1} 

Some visualization examples are provided for comparison:
    
    \begin{figure}[H]
        \centering
        \includegraphics[scale=0.5]{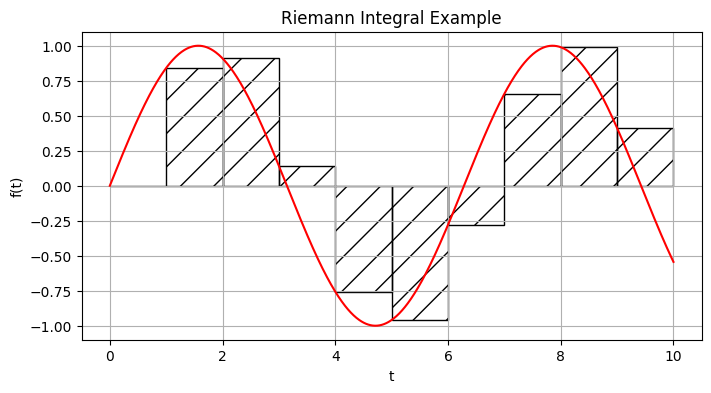}
        \caption{Riemann integral Example}
        \label{Riemann integral Examples}
    \end{figure}
    
    \begin{figure}[H]
        \centering
        \includegraphics[scale=0.5]{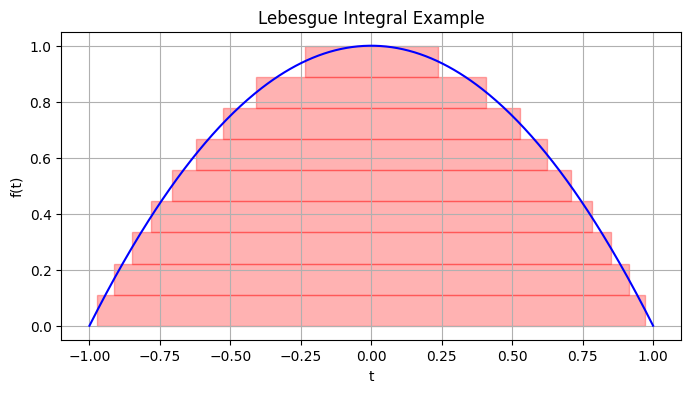}
        \caption{Lebesgue integral Example}
        \label{Lebesgue integral Example}
    \end{figure}

    \begin{figure}[H]
        \centering
        \includegraphics[scale=0.5]{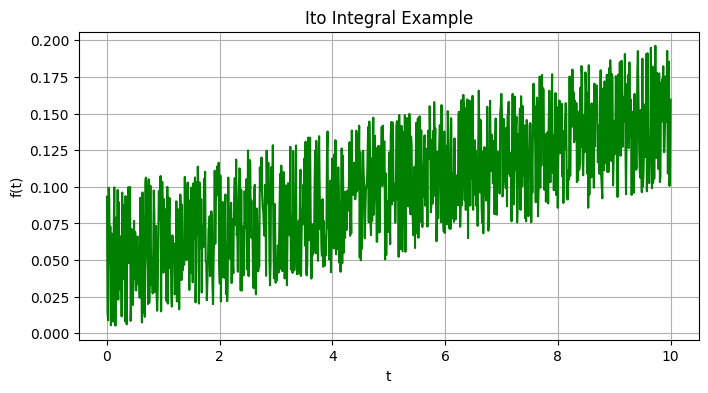}
        \caption{Itô integral Example}
        \label{Ito integral Example}
    \end{figure}

In finance, the \textbf{Riemann integral} is often used in deterministic models. For instance, the computation of the present value of a stream of cash flows employs the Riemann integral when the cash flow is continuously distributed over time. However, the Riemann integral falls short when we attempt to handle functions with high oscillations or discontinuities, which are common in financial markets.

The \textbf{Lebesgue integral} resolves many of the problems faced by the Riemann integral. It is capable of integrating a wider class of functions, making it more suitable for applications in probability theory and stochastic processes. Crucially, it allows for defining the expectation and variance of a random variable in a probability space, a concept pivotal in quantitative finance. For example, in the Black-Scholes-Merton model for option pricing, the risk-neutral expectation is used to determine the price of the option. This expectation is fundamentally a Lebesgue integral.

The \textbf{Itô integral} is a cornerstone in stochastic calculus, a branch of mathematics used extensively in mathematical finance. It is the most suitable for the stochastic calculus for finance of continuous-time models because it specifically takes into account the probabilistic aspects and is designed to handle stochastic processes. For instance, the Itô integral is employed to model the prices of financial derivatives.

\vspace{10mm}  

 In the context of derivative pricing, a partial differential equation (PDE) is an equation that relates the price of derivative security to the underlying asset's price and time. The PDE incorporates various factors such as the underlying asset's volatility, interest rates, and dividend yields to determine the fair value of the derivative. Solving the PDE provides a mathematical framework for pricing derivatives and determining their sensitivity to market factors.

Stochastic Differential Equations (SDEs) and Partial Differential Equations (PDEs) are interconnected in several ways, especially within the field of financial mathematics.
        
        \begin{definition}[\textbf{Stochastic Differential Equation (SDE)}]

            A stochastic differential equation is a differential equation that involves both deterministic and stochastic components. It is represented in the form
            \[
            dX(u) = \beta(u, X(u)) du + \gamma(u, X(u)) dW(u).
            \]
            
            Here, $\beta(u, x)$ represents the drift coefficient, and $\gamma(u, x)$ represents the diffusion coefficient. These coefficients are given functions that determine the behavior of the equation.
            
            The SDE is defined on a time interval $[0, T]$ and satisfies an initial condition $X(t) = x$, where $t \geq 0$ and $x \subseteq \mathbb{R}$.
    \end{definition}

The connection between SDEs and PDEs is prominently highlighted in the realm of options pricing, where the Black-Scholes-Merton model serves as a leading example. This model utilizes an SDE to describe the evolution of a stock price over time. Simultaneously, it employs a PDE, known as the Black-Scholes equation, to define the price of a derivative instrument, like an option, as a function of the stock price and time. More details about the Black-Scholes equation will be covered in the following subsection. 

With SDEs, additional financial products and market patterns can be analyzed mathematically. Before we dive into the interpretations and research section, a few terminologies need to be defined to give some real-world meaning to promote a clear picture of the concepts. 

    \begin{definition}[\textbf{Underlying Price of an Asset}]
        The underlying price of an asset refers to the current market value of the asset upon which a derivative security is based. It represents the price at which the asset can be bought or sold in the open market. The underlying asset could be a stock, bond, commodity, or any other instrument which can be traded.
    \end{definition}

    \begin{definition}[\textbf{Markov Derivative Security}]
         A Markov derivative security is a financial instrument whose value depends solely on the current state of the underlying asset and not on the historical path it took to reach that state. Markovian property implies that the derivative's future dynamics are determined by the current state of the underlying asset, making it a simpler and more tractable model for pricing and risk analysis.
    \end{definition}

    With PDEs and SDEs, finding the Pricing Differential Equation and Constructing a Hedge becomes feasible. The pricing differential equation is derived by considering the state process, which describes the dynamics of the underlying asset's price. By incorporating the state process into the PDE framework, one can derive a differential equation that captures the derivative security's pricing dynamics.
        
    To construct a hedge for a derivative security, one aims to offset the risk exposure of the derivative by taking an opposite position in the underlying asset or related instruments. This can involve adjusting the quantity of the underlying asset or incorporating other financial instruments to achieve a risk-neutral position. The hedge aims to reduce or eliminate the potential fluctuations in the derivative's value caused by changes in the underlying asset's price, thus mitigating risk.
        
    Another common terminology in real world is the discounted security price differential. It refers to the change in the derivative security's price when discounted at the risk-free rate. By discounting the future cash flows, the differential accounts for the time value of money.
    
    Under certain conditions, the discounted security price differential becomes a martingale under the risk-neutral measure. This means that its expected value at any time is equal to its current value, and it does not exhibit any predictable patterns or biases. The martingale property is crucial in options pricing and risk management, as it allows for the construction of replicating portfolios and the estimation of fair derivative prices using risk-neutral probabilities.
        
    The connection between these concepts lies in the development of pricing models for derivative securities. By considering the state process, incorporating discounted security price differentials, and analyzing when they become martingales under the risk-neutral measure, one can derive pricing equations and construct effective hedging strategies for managing derivative risks.

\subsubsection{Black-Scholes-Merton Model}\label{sec: BSmodel}

The Black-Scholes-Merton Model (the BSM model), proposed in Fischer Black and Myron Scholes's paper ``The Pricing of Options and Corporate Liabilities" (1973), serves as the cornerstone of contemporary financial theory. This model outlines a novel methodology for valuing various financial instruments, such as options, stocks, corporate bonds, and warrants, ingeniously incorporating parameters like the current stock price, the option's strike price, time to expiration, and the risk-free interest rate.

The following section introduces the Black-Scholes-Merton equation. For more detailed derivation and explanation, please refer to Chapter 4.5 and Chapter 5.2 of Shreve's ``Stochastic Calculus for Finance II: Continuous-Time Models"\cite {SCFII}.

A list of notations related to the BSM model is provided for reference.

\textbf{Notations:}
\begin{itemize}

\item $t$: the current time t,
\item $T$: the time of option expiration/maturity,
\item $\tau$: the time to expiration, $\tau = T - t$,
\item $S(t)$: the stock price at time t,
\item $x$: the current stock price, $x = S(t)$,
\item $K$: the strike/exercise price,
\item $(S(T) - K)^{+}$: the price for a European call option at time T,
\item $r$: the risk-free interest rate
\item $\sigma$: the stock volatility
\item $X(t)$: the portfolio valued at time t,
\item $X(0)$: the initial capital value
\item $W(t)$: the Brownian motion at time t,
\item $\Delta$: the number of shares of the stock in the portfolio,
\item $c(t,x)$: the value of call at time t, $c(t,x) = c(t, S(t))$,
\item $c_x(t, S(t))$: the delta of the option at time t,

\item $BSM(\tau,x;K,r,\sigma)$: the Black-Scholes-Merton Function,
\item $N(x)$: The standard normal cumulative distribution function (equation \ref{eq: The standard normal cumulative distribution function}),
\item $N'(x)$: The standard normal probability density function (equation \ref{eq: The standard normal probability density function}).
\end{itemize}

\begin{equation} \label{eq: The standard normal cumulative distribution function}
    N(x) = \frac{1}{\sqrt{2\pi}} \int_{-\infty}^{x} e^{-z^{2}/2} dz
\end{equation}
    
\begin{equation} \label{eq: The standard normal probability density function}
    N'(x) = \frac{dN(x)}{dx} = \frac{1}{\sqrt{2\pi}} e^{-x^{2}/2}
\end{equation}

\vspace{10mm}

Given a portfolio with value $X(t)$ at time t, it is invested in a money market with an interest rate $r$ under a stock model. The portfolio is composed of $\Delta(t)$ shares of stock at time $t$. Thus we have a total stock worth of the number of shares times the price per share of the stock, $\Delta(t)\cdot S(t)$, and obtain a cash position as the reminder, $X(t) - \Delta(t) \cdot S(t)$, which can be invested for interest earning and be represented as $r(X(t) - \Delta(t) \cdot S(t))$.

Therefore, we obtain the change of portfolio value $dX(t)$ as:

\begin{equation} \label{eq: change of portfolio value}
        dX(t) = \Delta(t) \cdot dS(t) + r(X(t) - \Delta(t) \cdot S(t))dt.
\end{equation}

Let $W(t)$, $0 \leq t \leq T$, be a Brownian motion on a probability space $(\Omega, \mathcal{F}, \mathbb{P})$, and let $\mathcal{F}(t)$, $0 \leq t \leq T$, be a filtration for this Brownian motion. We denote the stock price process as 

\begin{equation}
    dS(t) = \alpha(t) S(t)dt + \sigma(t) S(t)dW(t), 0 \leq t \leq T.
\end{equation}

We can modify the above equation by dividing $S(t)$ from both sides and get: 

\begin{equation}
    \frac{dS(t)}{S(t)} = \alpha(t) dt + \sigma(t) dW(t), 0 \leq t \leq T.
\end{equation}

The mean rate of return $\alpha(t)$ and the volatility $\sigma(t)$ are allowed to be adapted processes. For all $t \in [0, T]$, $\sigma(t)$ is assumed to be not zero almost surely.

Therefore we can expand equation \ref{eq: change of portfolio value} to:

\begin{equation}
    \begin{split}
        dX(t) &= \Delta(t) \cdot dS(t) + r(X(t) - \Delta(t) \cdot S(t))dt, \\
        &= - \Delta(t)(\alpha S(t)dt + \sigma S(t)dW(t)) + r(X(t) - \Delta(t)S(t))dt, \\
        &= rX(t)dt + \Delta(t)(\alpha -r)S(t)dt - \sigma \Delta(t)S(t)dW(t).
    \end{split}
\end{equation}

Let's denote the discounted stock price (Definition \ref{def: Risk-Neutral}) and the discounted portfolio value of an agent, $e^{-rt}S(t)$ and $e^{-rt}X(t)$, respectively. 

Applying the Itô-Doeblin formula with $f(t,x) = e^{-rt}x$, we have the differential of the discounted stock price as:

\begin{equation}
    d(e^{-rt}S(t)) = df(t, S(t)) = (\alpha - r)e^{-rt}S(t)dt + \sigma e^{-rt}S(t)dW(t),
\end{equation}

and the differential of the discounted portfolio value as:

\begin{equation} \label{eq: the differential of the discounted portfolio value }
    d(e^{-rt}X(t)) = df(t, X(t)) = \Delta(t)d(e^{-rt}S(t)).
\end{equation}

Equation \ref{eq: the differential of the discounted portfolio value } entails that the change in the discounted portfolio value depends solely on the change in the discounted stock price with respect to time $t$.

Consider a European call option that pays $(S(T) - K)^{+}$ at the maturity time $T$. This call value only depends on two variables: the time to expiration and the stock value at the time. It also depends on other parameters $r$ and $\sigma$. Therefore, we denote $c(t,x)$ as the call value at time t, given at the time stock price is $S(t) = x$. The function $c(t,x)$ is not random; however, the option value is random: it is the stochastic process $c(t, S(t))$ generated by substituting $x$ with the random stock price $S(t)$. 

The nature of the stock market is that the future stock price is random. Thus, the future option price $c(t, S(t))$ is also unknown to people. The goal here is to determine the function $c(t,x)$, such that we can produce a formula for future option value, given future stock price.

We begin with computing the differentials for both $c(t, S(t))$ and the discounted option price $e^{-rt}c(t, S(t))$ using the Itô-Doeblin formula:

\begin{equation} \label{eq: 4.5.6}
\begin{split}
    dc(t, S(t)) = & c_t(t, S(t))dt + c_x(t, S(t))dS(t) + \frac{1}{2}c_{xx}(t,S(t))(dS(t))^2 \\
    = & [c_t(t, S(t)) + \alpha S(t) c_x(t, S(t)) + \frac{1}{2}\sigma^2S^2(t)c_{xx}(t,S(t))]dt \\
    & + \sigma S(t)c_x(t,S(t))dW(t) 
\end{split}
\end{equation}

and, let $f(t,x) = e^{-rt}x$,

\begin{equation} \label{eq: 4.5.7}
\begin{split}
    d(e^{-rt}c(t, S(t))) = & df(t, c(t, S(t))) \\
    = & e^{-rt}[-rc_t(t, S(t)) + c_t(t, S(t)) + \alpha S(t) c_x(t, S(t)) \\
    & + \frac{1}{2}\sigma^2S^2(t)c_{xx}(t,S(t))]dt + e^{-rt}\sigma S(t)c_x(t,S(t))dW(t).
\end{split}
\end{equation}

A (short option) hedging portfolio starts with $X(0)$, the initial capital. It invests in both the money market (e.g. bank savings) and stock account, thus $X(t)$, the portfolio value at time $t \in [0, T]$ agrees with $c(t, S(t))$. This holds if and only if 

\begin{equation}
     e^{-rt}X(t) = e^{-rt}c(t, S(t)), \quad \forall t \in [0, T),
\end{equation}

under the condition:

\begin{equation} 
    X(0) = c(0, S(0)).
\end{equation}

In particular: 

\begin{equation}
     e^{-rt}X(t) - X(0) = e^{-rt}c(t, S(t)) - c(0, S(0)), \quad \forall  t \in [0, T).
\end{equation}

Recall the differential of the discounted portfolio value from equation \ref{eq: the differential of the discounted portfolio value }, comparing equation \ref{eq: the differential of the discounted portfolio value } and equation \ref{eq: 4.5.7}, equation

\begin{equation}
    d( e^{-rt}X(t)) = d(e^{-rt}c(t, S(t))), \quad \forall t \in [0, T)
\end{equation}

happens if and only if:

\begin{equation} \label{eq: 4.5.10}
\begin{split}
    & \Delta(t)(\alpha - r)S(t)dt + \Delta(t)\sigma S(t)dW(t) \\
    & = [-rc_t(t, S(t)) + c_t(t, S(t)) + \alpha S(t) c_x(t, S(t))+ \frac{1}{2}\sigma^2S^2(t)c_{xx}(t,S(t))]dt                \\
    & + \sigma S(t)c_x(t,S(t))dW(t).
\end{split}
\end{equation}

Equating the $dW(t)$ term returns the delta-hedging rule:

\begin{equation} \label{eq: 4.5.11}
    \Delta(t) = c_x(t, S(t)), \quad \forall  \in [0, T).
\end{equation}

$c_x(t, S(t))$ here refers to the $\mathcal(delta)$ of the option.

Equating the $dt$ term with equation \ref{eq: 4.5.10}, we can obtain:

\begin{equation} \label{eq: 4.5.12}
 \begin{split}
    & (\alpha - r)S(t) + \sigma S(t)dW(t) \\
    & = -rc_t(t, S(t)) + c_t(t, S(t)) + \alpha S(t) c_x(t, S(t))+ \frac{1}{2}\sigma^2S^2(t)c_{xx}(t,S(t))                \\
    & + \sigma S(t)c_x(t,S(t)), \quad \forall  \in [0, T).
\end{split}   
\end{equation}

Canceling $\alpha S(t)c_x(t,S(t))$ from both sides of equation \ref{eq: 4.5.12} returns:

\begin{equation} \label{eq: 4.5.13}
    rc(t,S(t) = c_t(t,S(t)) + rS(t)c_x(t,S(t)) + \frac{1}{2}\sigma^2S^2(t)c_{xx}{t,S(t)}, \quad \forall  \in [0, T).
\end{equation}

As a result, if setting $x = S(t)$, we want to find the solution, $c(t,x)$, to \textbf{the Black-Scholes-Merton partial differential equation}: (Notice: \ref{notice: solution})

\begin{equation} \label{eq: Black-Scholes-Merton equation}
    c_t(t,x) + rxc_x(t,x) + \frac{1}{2}\sigma^2x^2(t)c_{xx}{t,x} = rc(t,x)
\end{equation}
    for all $t\in [0,T), x\geq 0$,  and also satisfies the \textbf{terminal condition}:
    
    \begin{equation}\label{eq: terminal condition}
        c(T,x) = (x - K)^{+}.
    \end{equation}

Equation \ref{eq: Black-Scholes-Merton equation} is a type of backward parabolic. We need boundary conditions at $x = 0$ and $x = \infty$ to determine the solution, in addition to the terminal condition \ref{eq: terminal condition}.

Substituting $x = 0$ into equation \ref{eq: Black-Scholes-Merton equation} gives an ordinary differential equation (ODE):

\begin{equation}
    c_t(t,0) = rc(t,0),
\end{equation}

with solution:

\begin{equation}
    c(t,0) = e^{rt}c(0,0).
\end{equation}

Additionally, if we substitute $t = T$ and use $c(T,0) = (0 - K)^{+} = 0$,
we can get $c(0,0) = 0$ and therefore generates the \textbf{boundary condition at $x = 0$}

\begin{equation} \label{eq: boundary conditions 1}
    c(t,0) = 0 
\end{equation}
for all $t \in [0,T]$.

For the case $x \rightarrow \infty$, $c(t,x)$ increases without converging. Thus we specify the rate of growth for the boundary condition at $x = \infty$. A \textbf{boundary condition at $x = \infty$}, specified by Shreve\cite{SCFII}, for the European call is:

\begin{equation} \label{eq: boundary conditions 2}
    \lim_{x \rightarrow -\infty}[c(t,x) - (x - e^{-r(T-t)}K)] = 0
\end{equation}

for all $t \in [0, T]$.

    The solution to the Black-Scholes-Merton (BSM) equation \ref{eq: Black-Scholes-Merton equation} with terminal condition \ref{eq: terminal condition} and boundary conditions \ref{eq: boundary conditions 1} and \ref{eq: boundary conditions 2} is: 
    
    \begin{equation}
            c(t,x) = xN(d_{+}(T-t, x)) - Ke^{-r(T-t)}N(d_{-}(T-t,x)), 
    \end{equation}
    given $0\leq t \leq T, x > 0$,
    
    where

    \begin{equation}
        d_\pm(\tau, x) = \frac{1}{\sigma\sqrt{\tau}}\left[\log\left(\frac{x}{K}\right) + \left(r \pm \frac{\sigma^2}{2}\right)\tau\right],
    \end{equation}
    
    and N is the cumulative distribution function for the standard normal distribution:
    
    \begin{equation}
        N(y) =  \frac{1}{\sqrt{2\pi}} \int_{-\infty}^{y} e^{-z^{2}/2} dz.
    \end{equation}

    The Black-Scholes-Merton (BSM) function is sometimes denoted as:

    \begin{equation} \label{eq: 4.5.22}
            BSM(\tau,x;K,r,\sigma) = xN(d_{+}(\tau, x)) - Ke^{-r\tau}N(d_{-}(\tau, x)).
    \end{equation}

In this section, we derive the Black-Scholes-Merton PDE \ref{eq: Black-Scholes-Merton equation} and then provide the solution in equation (4.5.19) without explaining how the solution is obtained. In section \ref{Risk-Neutral Valuation for Deriving the BSM Formula}, we will show how to derive the solution under the risk-neutral (equivalent martingale) measure.\label{notice: solution}

\subsubsection{The Greeks}

Previously, the Black-Scholes-Merton model is introduced for pricing options, assuming that the underlying asset price follows a geometric Brownian motion. The \textbf{Greeks}, on the other hand, are measures describing how the option price changes with respect to different impacts. They are derived from the Black-Scholes model and help traders and investors understand and manage the risks associated with options positions.

The table lists a brief explanation of the connection between the Black-Scholes model and the Greeks:

    \begin{itemize}
        \item \textbf{Delta ($\Delta$)}: Delta measures the sensitivity of an option's price to changes in the price of the underlying asset. It represents the rate of change of the option price with respect to the underlying asset price.
        
        \item \textbf{Gamma ($\Gamma$)}: Gamma represents the rate of change of an option's delta in response to changes in the price of the underlying asset. It measures the curvature of the option's price curve.
        
        \item \textbf{Theta ($\Theta$)}: Theta measures the rate of decline in the value of an option over time as the expiration date approaches. It captures the effect of time decay on the option's price.
        
        \item \textbf{Vega ($\nu$)}: Vega measures the sensitivity of an option's price to changes in implied volatility. It quantifies the impact of changes in market expectations of future volatility on the option price.
        
        \item \textbf{Rho ($\rho$)}: Rho measures the sensitivity of an option's price to changes in the risk-free interest rate. It represents the rate of change of the option price with respect to changes in the risk-free interest rate.
        
    \end{itemize}

    The understanding of these Greek letters is crucial for managing and evaluating options and derivatives strategies, as they allow investors to assess the potential risks and rewards associated with different market conditions and price movements.
    
    For example, \textbf{Delta-Neutral} refers to a portfolio or position in which the total delta is zero. Delta measures the sensitivity of the option or portfolio value to changes in the underlying asset price. By creating a delta-neutral position, investors aim to eliminate the directional risk associated with the underlying asset's price movements, focusing instead on other sources of potential profit or loss. To illustrate a \textbf{Delta-Neutral Position}, let's consider an investor with a portfolio consisting of 100 call options on a sample stock, each with a delta of 0.6, and simultaneously short-sells 60 shares of the same stock, each with a delta of -1.0.

The \textbf{portfolio delta} ($\Delta_{Portfolio}$) can be calculated using the formula:

\begin{equation}
    \label{eq: Portfolio Delta}
    \begin{aligned}
        \Delta_{Portfolio} &= 
        & (\text{\# Calls} \times \text{$\Delta$ per Call})
        & + (\text{\# Shares} \times \text{$\Delta$ per Share}).
    \end{aligned}
\end{equation}

In this example, the portfolio delta is:

\[
\Delta_{Portfolio} = (100 \times 0.6) + (-60 \times -1.0) = 60.
\]

To achieve a delta-neutral position, the investor would need to adjust the portfolio by selling 60 additional call options, each with a delta of -0.6. This action effectively reduces the portfolio delta to zero.

By achieving a delta-neutral position, investors eliminate the exposure to the directional movement of the underlying stock. The profitability of the position will depend on other factors, such as changes in implied volatility, time decay (theta), and the stock's price reaching certain levels, rather than its general direction.

\subsubsection{Risk-Neutral Measure}\label{sec: Risk-Neutral Measure}

The field of financial mathematics frequently uses the technique of change of measure, particularly in the valuation of financial derivatives. Under the equivalent martingale measure or the \textbf{risk-neutral measure} (the risk-free measure), the discounted price process of a tradable asset is a martingale. This fundamental concept underlies the risk-neutral valuation method. To simplify the pricing of derivatives, the change of measure technique is used to shift from the actual probability measure to the risk-neutral measure.

With Girsanov's Theorem, we can update the binomial asset pricing model for a single underlying security for the continuous models, and how we program for multiple underlying securities with constraints of hedging for non-arbitrages. Girsanov's Theorems are a set of results in stochastic analysis that establish a connection between change of measure, stochastic processes, and the concept of martingales. These theorems provide a powerful framework for analyzing and transforming stochastic processes under different measures.

\begin{theorem}[\textbf{Girsanov, multi-dimensions}]\label{thm: Girsanov}
    Let T be a fixed positive time, and let \\
    $\Theta(t) = (\Theta_1(t), \dots, \Theta_d(t))$ 
    be a d-dimensional adapted process. Define: \\
    
    \begin{equation} 
    Z(t) = exp\{ -\int_0^t \Theta(u)dW(u) - \frac{1}{2}\int_0^t \Theta^2(u)du)\},
    \end{equation}

    \begin{equation}
            \tilde{W}(t) = W(t) + \int_0^t \Theta(u)du.
    \end{equation}

    If \textbf{Novikov's condition},

    \begin{equation}
        E[exp{\frac{1}{2}\int_0^T |\Theta(u)|^2dt}] < \infty,
    \end{equation}

    is satisfied, then the process,

    \begin{equation}
        \epsilon(\int_0^t \Theta(u)dW(u)) = exp({\int_0^T \Theta(u)dW(t)-\frac{1}{2}\int_0^t |\Theta(u)|^2dt}) , 0 \leq t \leq T,
    \end{equation}
    
    is a martingale under the probability measure $\mathbb{P}$  and the filtration $\mathcal{F}$. The $\epsilon$ here represents the Doléans-Dade exponential.

\end{theorem}

    \begin{definition}[\textbf{the Doléans-Dade exponential}]\label{the Doléans-Dade exponential}

        If the semi-martingale X is continuous, then the Doléans-Dade exponential or the stochastic exponential of X is defined as:

        \begin{equation}
            \epsilon(X) = exp(X-X_0-\frac{1}{2}[X]).
        \end{equation}
        
        In addition, if X is a Brownian motion, then the Doléans-Dade exponential is a geometric Brownian motion.
        
    \end{definition}

Consider a financial market consisting of a risk-free stock. The stock price undergoes changes according to a geometric Brownian motion under the real-world measure $\mathbb{P}$:

\[
dS_t = \mu S_t dt + \sigma S_t dW_t^\mathbb{P},
\]

where $W_t^\mathbb{P}$ represents a $\mathbb{P}$-Brownian motion. By utilizing Girsanov's theorem, we can define a $\mathbb{Q}$-Brownian motion $W_t^\mathbb{Q}$ such that

\[
dW_t^\mathbb{Q} = dW_t^\mathbb{P} + \lambda dt,
\]

where $\lambda = \frac{\mu - r}{\sigma}$. By the change of measure, the dynamics of the stock price under the risk-neutral measure $\mathbb{P}$ is:

\[
dS_t = r S_t dt + \sigma S_t dW_t^\mathbb{P},
\]

This change of measure from $\mathbb{P}$ to $\mathbb{P}$ simplifies the pricing of derivatives by permitting us to discount expected future payoffs at the risk-free rate $r$. Thus we formally define the risk-neutral probably measure:

\begin{definition}[\textbf{Risk-Neutral}] \label{def: Risk-Neutral}
    A probability measure $\tilde{\mathbb{P}}$ is said to be risk-neutral if: \\
    (i) $\tilde{\mathbb{P}}$ and $\mathbb{P}$ are equivalent (i.e., for every $A \in \mathcal{F}, \mathbb{P}(A) = 0$) if and only if  $\tilde{\mathbb{P}}(A) = 0$, and  \\
    (ii) under $\tilde{\mathbb{P}}$, the discounted stock price $D(t)S_i(t)$ is a martingale for every $i = 1,\dots,m$.
\end{definition}

\begin{definition}[\textbf{Discount Process}] \label{def: Discount Process}
    The discount process is defined as

    \begin{equation}
        D(T) = exp(-\int_0^t R(s)ds),
    \end{equation}

    where $R(t)$ refers to an adapted interest rate process.

    Define $I(t) = \int_0^t R(s)ds$, we can obtain $dI(t) = R(t)dt$ and $dI(t)dI(t) = 0$. Thus we can compute $dD(t)$ using the Itô-Doeblin formula:
    
    \begin{equation}
       d D(T) = df(I(t)) = -R(t)D(t)dt.
    \end{equation}
    
\end{definition}

Notice, D(t) has zero quadratic variation.

\begin{theorem} \label{thm: discounted portfolio value is a martingale}
    Let $\tilde{\mathbb{P}}$ be a risk-neutral measure, and let $X(t)$ be the value of a portfolio. Under $\tilde{\mathbb{P}}$, the discounted portfolio value $D(t)X(t)$ is a martingale. 
\end{theorem}
    
Theorem \ref{thm: discounted portfolio value is a martingale} is further discussed and applied in the context of the Martingale Representation Theorem and the derivation of the Black-Scholes-Merton Model.

In financial mathematics, the \textbf{risk-neutral probability} is a measure of probability used to price derivative securities. It is a hypothetical probability measure under which the expected return on an asset is equal to the risk-free interest rate. By using the risk-neutral probability, we can value derivatives without considering the market participants' risk preferences.

Let $\tilde{\mathbb{P}}$ be the risk-neutral probability measure and $S(t)$ be the price of a stock at time $t$. The risk-neutral probability measure is defined such that the discounted stock prices are martingales under this measure.

Under the risk-neutral probability measure $\tilde{\mathbb{P}}$, the discounted stock prices follow a martingale process. Mathematically, this can be represented as:

\begin{equation}
    \frac{{dS(t)}}{{S(t)}} = r \cdot dt + \sigma \cdot \tilde{W}(t).
\end{equation}

The above equation implies that the expected rate of return on the stock price is equal to the risk-free interest rate $r$. This assumption allows us to value derivatives using the risk-neutral probability measure, as it simplifies the pricing process by removing considerations of risk preferences.

For an undiscounted stock price $S(t)$, its mean rate of return is equal to the interest rate under the risk-neutral measure $\tilde{\mathbb{P}}$. The formula is given as:

\begin{equation} \label{eq: 5.2.24}
    S(t) = S(0)exp\{ \int_0^t \sigma(s)d\tilde{W}(s) + \int_0^t(R(s) - \frac{1}{2}\sigma^2(s))ds    \}.
\end{equation}

By considering the discounted stock prices as martingales under the risk-neutral probability measure, we can apply techniques such as the Black-Scholes formula to price options and other derivative securities.

\subsubsection{Martingale Representation Theorem}

In financial models, assumptions such as non-arbitrage and risk-neutral conditions are often made. Complex factors like dividends, transaction costs, and inflation are usually omitted. Under these simplifications, we can derive risk-neutral pricing results, which heavily rely on the \textbf{Martingale Representation Theorem} in continuous models\cite{Elliott} \cite{SCFII}. This section presents the basics of the Martingale Representation Theorem. Under the risk-neutral measure, its application of replicating options through the Black-Scholes-Merton model to price European options is presented in the following section \ref{sec: BSmodel}.

\begin{theorem}[\textbf{Martingale Representation Theorem, one dimension}] \label{thm:martingale_representation}

    Let $W_{t}$, $0 \leq t \leq T$, be a Brownian motion on a probability space $(\Omega, \mathcal{F}, P)$, and let $\mathcal{F}(t)$, $0 \leq t \leq T$, be the filtration generated by this Brownian motion. Let $M(t)$, $0 \leq t \leq T$, be a martingale with respect to this filtration (i.e., for every t, $M(t)$ is $\mathcal{F}(t)$-measurable and for $0 \leq s \leq t \leq T$, $E[M(t) |\mathcal{F}(s)] = M(s)$). Then there is an adapted process $\Gamma(u)$, $0 \leq u \leq T$, such that: 

    \begin{equation}
    M(t) = M(0) + \int_{0}^{t} \Gamma(u) dW(u), \quad 0 \leq t \leq T.
    \end{equation}
    
\end{theorem}

The one-dimension Martingale Representation Theorem, Theorem \ref{thm:martingale_representation}, assumes the filtration is generated by the Brownian motion. This is more restrictive than the assumption from Theorem \ref{thm: Girsanov}. Thus we can update Girsanov Theorem:

\begin{theorem}
    Let $\tilde{M}(t)$, $0 \leq t \leq T$, be a martingale under $\tilde{\mathbb{P}}$. Then there is an adapted process $\tilde{\Gamma}(u)$, $0 \leq t \leq T$, such that

    $\tilde{M}(t) = \tilde{M}(0) + \int_0^t \tilde{\Gamma}(u)d\tilde{W}(u)$, $0 \leq t \leq T$.
    
\end{theorem}

If there exist multiple underlying securities in the portfolio, one may use the multi-dimensional Martingale Representation Theorem to replicate the products.

\begin{theorem}[\textbf{Martingale Representation Theorem, multi-dimensions}] \label{thm:martingale_representation 2}

    Let T be a fixed positive time, and assume that $\mathcal{F}(t)$, $0 \leq t \leq T$, is the filtration generated by the d-dimensional Brownian motion $W_{t}$, $0 \leq t \leq T$, 
    
    Let $M(t)$, $0 \leq t \leq T$, be a martingale with respect to this filtration under $\mathbb{P}$. Then there is an adapted, d-dimensional process $\Gamma(u) = (\Gamma_1(u), \dots, \Gamma_d(u))$, $0 \leq u \leq T$, such that: 

    \begin{equation}
        M(t) = M(0) + \int_{0}^{t} \Gamma(u) dW(u), \quad 0 \leq t \leq T.
    \end{equation}

    If, in addition, we assume the notation and assumptions of Theorem \ref{thm: Girsanov} and if $\tilde{M(t)}$,  $0 \leq t \leq T$, is a $\tilde{\mathbb{P}}$-martingale, then there is an adapted, d-dimensional process $\tilde{\Gamma}(u) = (\tilde{\Gamma_1}(u)), \dots, \tilde{\Gamma_d}(u)$ such that

    \begin{equation}
        \tilde{M}(t) = \tilde{M}(0) + \int_0^t \tilde{\Gamma(u)}d\tilde{W}(u), 0 \leq t \leq T.
    \end{equation}

\end{theorem}

\vspace{10mm}

Fundamental Theorems of Asset Pricing are a set of key results in mathematical finance that establish the relationship between the absence of arbitrage opportunities and the existence of an equivalent martingale measure.

\begin{theorem} \textbf{Fundamental Theorems of Asset Pricing} \label{thm: Fundamental Theorems of Asset Pricing}

    \begin{enumerate}

        \item   \textbf{First Fundamental Theorem of Asset Pricing}: 
            
            Let $(\Omega, \mathcal{F}, (\mathcal{F}_t)_{t \geq 0}, P)$ be a filtered probability space representing the financial market. Assume that there are no arbitrage opportunities in the market. Then, there exists an equivalent martingale measure $Q$ such that the discounted asset prices are martingales under $Q$.
    
        \item   \textbf{Second Fundamental Theorem of Asset Pricing}: 

             Let $(\Omega, \mathcal{F}, (\mathcal{F}_t)_{t \geq 0}, P)$ be a filtered probability space representing the financial market. Assume that the market is complete, meaning that every contingent claim can be perfectly hedged. Then, there exists a unique equivalent martingale measure $Q$ that prices all contingent claims in the market.
    \end{enumerate}

\end{theorem}

        The fundamental theorems of asset pricing provide fundamental insights into the relationship between the absence of arbitrage opportunities, the existence of equivalent martingale measures, and the pricing of contingent claims in a financial market. These theorems serve as cornerstones in mathematical finance and have profound implications for pricing derivatives, risk management, and the valuation of financial assets.

The Martingale Representation Theorem and the Fundamental Theorems of Asset Pricing form the basis for the Black-Scholes-Merton model under risk-neutral measure, used for pricing European call and put options. In this model, the price process of the risky asset follows a geometric Brownian motion and hence is a martingale under the risk-neutral measure. Thus, a derivative security that depends on the asset's price at maturity can be replicated and consequently priced through continuous trading in the risky asset.

\subsubsection{Risk-Neutral Valuation for Deriving the BSM Formula} \label{Risk-Neutral Valuation for Deriving the BSM Formula}

In section \ref{sec: BSmodel}, we introduce the Black-Scholes-Merton equation for a European call option. The change from the actual measure $\mathbb{P}$ to the risk-neutral measure $\tilde{\mathbb{P}}$ affects the stock's mean rate of return, but the volatility stays constant.

Following the risk-neutral discussion from section \ref{sec: Risk-Neutral Measure}, we set $X_0$, the initial capital, and $\Delta(t)$, the portfolio process for an agent to hedge a short position in the call (which achieves $X(T) = (S(t) - K)^{+}$ almost surely). We hereby generalize the equation for pricing under the risk-neutral measure.

Let $V(T)$, the payoff at time $T$ of a derivative security, be an $\mathcal{F}(T)$-measurable random variable. The payoff at time $T$ of the derivative security is path-dependent. Considering that an agent seeks to hedge a short position for an underlying security, we need to find $X(0)$, the initial capital, and $\Delta(t)$, the portfolio process, given $0 \leq t \leq T$.

This scenario can be described find $V(T)$ such that 
    \begin{equation} \label{eq: 5.2.28}
        X(T) = V(T).
    \end{equation}
    
holds almost surely, without assuming the mean return rate, volatility, and interest rate to be constant. The process of choosing $X(0)$ and $V(T)$ to satisfy equation \ref{eq: 5.2.28} can be achieved using the Martingale Representation Theorem. Upon completion, the discounted portfolio value $D(t)X(t)$ is a martingale under $\tilde{\mathbb{P}}$ as mentioned in theorem \ref{thm: discounted portfolio value is a martingale}. This result indicates that:

\begin{equation} \label{eq: 5.2.29}
    D(t)X(t) = \tilde{\mathbb{E}}[D(T)X(T)|\mathcal{F}_t] = \tilde{\mathbb(E)}[D(T)V(T)|\mathcal{F}_t].
\end{equation}

In this portfolio hedging case, $X(t)$ is the capital needed at time t to successfully hedge the short option with payoff $V(T)$. The risk-neutral pricing formulas for the continuous-time model are derived as the following:

\begin{equation} \label{eq: 5.2.30}
    D(t)V(t) = \tilde{\mathbb{E}}[D(T)V(T)|\mathcal{F}_t], 0 \leq t \leq T.
\end{equation}

\begin{equation} \label{eq: 5.2.31}
    V(t) = \tilde{\mathbb{E}}[exp(-\int_t^TR(u)du)V(T)|\mathcal{F}_t], 0 \leq t \leq T.
\end{equation}

These risk-neutral pricing formulas are widely applied in option pricing models. 

Continuing from risk-neutral pricing equation \ref{eq: 5.2.31}, the initial capital $V(0)$ is given as:

\begin{equation}
    V(0) = \tilde{\mathbb{E}}[exp(-\int_t^TR(u)du)V(T)|\mathcal{F}_t],
\end{equation}

assuming a portfolio process $\Delta(t)$ exists given an agent starts with the correct $X(0)$ and the portfolio will be $V(T)$ almost surely at the final time $T$. An example of how the one-dimensional Martingale Representation Theorem is applied to prove the assumption can be found in Chapter 5.3 of Shreve's ``Stochastic Calculus for Finance II: Continuous-Time Models"\cite{SCFII}.

To obtain the call price of a European option from the Black-Scholes-Merton model, we assume $\sigma$ and $r$ to be constant volatility and interest rate, respectively, and set $V(T) = (S(T) - K)^{+}$ as the payoff of the derivative security. We obtain a modified right-hand side of the equation \ref{eq: 5.2.31}:

\begin{equation} \label{eq: 5.2.31 modified}
     \tilde{\mathbb{E}}[exp(-r(T-t)(S(T) - K)^{+}|\mathcal{F}].
\end{equation}

Recall geometric Brownian motion is a Markov process, there exists a function $c(t,x)$ such that:

\begin{equation} \label{eq: 5.2.32}
    c(t,x) = \tilde{\mathbb{E}}[exp(-r(T-t)(S(T) - K)^{+}|S_t = x].
\end{equation}

Using the Independence Lemma based on the Properties of Conditional Expectation \ref{Properties of Conditional Expectation}, $c(t,x)$ can be computed, with constant $\sigma$ and $r$, as an update to equation \ref{eq: 5.2.24}:

\begin{equation} \label{eq: 5.2.24 modified}
    S(t) = S(0)exp\{\sigma \tilde{W}(t) + (r - \frac{1}{2}\sigma^2)t \}.
\end{equation}

Let's denote the time to expiration/maturity $\tau = T - t$ and the standard normal random variable $Y$ as:

\begin{equation} \label{the standard normal random variable $Y$}
    Y = -\frac{\tilde{W}(T)-\tilde{W}(t)}{\sqrt{T-t}}.
\end{equation}

We can define $S(T)$ based on equation \ref{eq: 5.2.24 modified} as:

\begin{equation}
\begin{split}
    S(T) & =  S(t)exp\{\sigma( \tilde{W}(T) - \tilde{W}(t)) + (r - \frac{1}{2}\sigma^2)\tau \}              \\
         & = S(t)exp\{-\sigma(\sqrt{\tau}Y) + (r - \frac{1}{2}\sigma^2)\tau \}. 
\end{split}
\end{equation}

$S(T)$ turns out to be the product of $S(t)$, the $\mathcal{F}$-measurable random variable, and the random variable 

\begin{equation}
    exp\{-\sigma(\sqrt{\tau}Y) + (r - \frac{1}{2}\sigma^2)\tau \},
\end{equation}

which is independent of $\mathcal{F}$.

As a result, equation \ref{eq: 5.2.32} stands with $c(t,x) = BSM(\tau,x;K,r,\sigma)$

\begin{equation}
    \begin{split}
            c(t,x) & = \tilde{\mathbb{E}}[e^{-r\tau}(xexp\{-\sigma \sqrt{\tau}Y + (r - \frac{1}{2}\sigma^2)  \}  -K)^{+}] \\
            & = \dots \label{skip parts} \\
            & = N(d_{+}(\tau, x)) - Ke^{-r\tau}N(d_{-}(\tau, x)).
    \end{split}
\end{equation}

For a complete deriving process forthe ``skipped $\dots$ parts" \ref{skip parts}, please refer to pages 219-220 in ``Stochastic Calculus for Finance II: Continuous-Time Models" \cite{SCFII}.

Therefore, we have shown 

\begin{equation} \label{eq: 5.2.36}
            BSM(\tau,x;K,r,\sigma) = xN(d_{+}(\tau, x)) - Ke^{-r\tau}N(d_{-}(\tau, x)),
    \end{equation}

which is identical to equation \ref{eq: 4.5.22}. The boundary and terminal conditions and the equations for $N(d_{\pm}(\tau, x))$ are also introduced in section \ref{sec: BSmodel}. 

\vspace{10mm}

Let's review the key ideas on Black-Scholes-Merton model.  In addition to the derivation using stochastic differential equations (SDE) and partial differential equations (PDE), we introduce the risk-neutral valuation approach to derive the BSM equation. This model is primarily based on the Geometric Brownian motion model (GBM).

Considering 

\begin{align*}
S_t &= S_0 e^{(\mu - \frac{1}{2} \sigma^2)t + \sigma B_t}
\end{align*}

the GBM is a martingale if and only if $\mu = 0$ and the discounted stock price is a martingale if and only if $\mu = r$. Hence, the BSm model is complete and has no arbitrage opportunities.

\subsection{Section Summary}

In this section, we embark on a journey of stochastic calculus for finance. We begin with the binomial asset pricing model, which simplifies the pricing of derivatives by dividing time into discrete intervals. We then transition to the continuous-time model, introducing stochastic calculus and its use in capturing the continuous and unpredictable nature of financial markets. The highlight of this section is the Black-Scholes model for option pricing under the risk-neutral measure.

The risk-neutral valuation for deriving the BSM Formula presents an example of applying a change of measure to replicate the portfolio. The desire to replicate the payout leads to making the portfolio a martingale, which enables the implementation of an investment strategy based on the Martingale Representation Theorem (Theorem \ref{thm:martingale_representation}). Under the risk-neutral measure, the initial cost to set up the portfolio is the price of the option. If the cost is below or above, there will be arbitrage in which investors can profit directly without taking risks. The arbitrage condition is not considered in this thesis.

Additional topics, such as Exotic Options, Change of Numéraire, and Term-Structure Models, and more detailed proofs and examples can be found in Shreve's ``Stochastic Calculus for Finance II".

Continuing our exploration of stochastic calculus for finance, we now shift our attention to the work of Peter Carr and his colleagues. By delving into Carr's research, we aim to gain deeper insights into the advancements and practical applications of stochastic calculus in finance.

\newpage
\section{\large Convex Duality in Continuous Option Pricing Models} \label{sec: ConvexDuality}

This thesis presents a novel approach by Carr and Torreicelli to understanding diffusive asset pricing models by utilizing the convex duality theory. The main focus is the paper ``Convex Duality in Continuous Option Pricing Models" by Peter Carr and Lorenzo Torricelli. In their paper, Carr and Torricelli propose an alternative approach to traditional asset pricing models by directly specifying a stochastic differential equation for the dual delta, which represents the option's convex conjugate (Legendre transform). This approach enables the derivation of option prices through the inversion of the Legendre transforms, satisfying an initial value problem (IVP) dual to the Dupire equation. \cite{ConvexDuality}

The authors first propose that the dynamics of the stock price are governed by a stochastic differential equation (SDE) in equation \ref{eq: 2.1} with the option value function from equation \ref{eq: 2.3} defined using the fundamental asset pricing theorem. Standard asset pricing theory mainly focuses on two types of PDEs for option pricing: the backward Kolmogorov equation (KBE) and the Dupire IVP. 

Modified from these PDEs, Carr and Torricelli state the conditions the option price should obey, which are derived from the probabilistic representation in equation \ref{eq: 2.3} and the stock price generator defined in equation \ref{eq: 2.1}. They introduce the adjoint processes in equations \ref{eq: 2.16}, \ref{eq: 2.17}, and definition \ref{definition 1}, and the relation between $\Delta$ and $P$ is given in Proposition \ref{proposition 1}. 

They present a change of measure from a risk-neutral probability measure to the dual delta measure in equation \ref{eq: 2.26}, proposition \ref{proposition 2}, and definition \ref{definition 2}. They prove the put option value as a solution to the Dual Dupire IVP in proposition \ref{proposition 3}. 

Lastly, Carr and Torricelli introduce a variable cross-sectional volatility model: the logistic model, and compare it with the normal model.

\subsection{Model Background}

Carr and Torricelli's research involves the development of mathematical models for continuous-time option pricing. This section first briefly reviews some of the founding mathematical concepts involved in ``Convex Duality in Continuous Option Pricing Models", then reviews how Carr and Torricelli develop their model in their paper.

In a chronological progression of mathematical models for option pricing mentioned in ``Convex Duality in Continuous Option Pricing Models", we begin with the Bachelier model presented in Louis Bachelier's Ph.D. thesis in 1900 \cite{bachelier1900}. Let's denote $F_t$ as the $T$-forward price of an asset at time $t$, $K$ as the strike price, and $T$ as the time-to-maturity \cite{guideBachelier}. 

Under the Bachelier model, the undiscounted price of a call option is 

\begin{equation}
    C_N(K) = (F_0 - K)N(d_N) + \sigma_N\sqrt{T}n(d_N),
\end{equation}

and the put option price is

\begin{equation}
    P_N(K) = (K - F_0)N(-d_N) + \sigma_N\sqrt{T}n(d_N),
\end{equation}

for

\begin{equation}
    d_N = \frac{F_0 - K}{\sigma_N\sqrt{T}},
\end{equation}

where $n(z)$ and $N(z)$ are the PDF and CDF, respectively, of the standard normal distribution.

The Bachelier model assumes that $F_t$ follows an arithmetic BM with volatility $\sigma_N$:

\begin{equation}    
         F_t = F_0 + \sigma_N dW_t,
\end{equation}

where $W_t$ is a standard BM under the $T$-forward measure.

In differential form, if we denote $S_t = F_t$ and $\sigma(S_t,t) = \sigma_N$, the Bachelier model follows:

\begin{equation}    
         dS_t = \sigma(S_t,t) dW_t.
\end{equation}

Subsequent advancements were made by Black, Scholes, and Merton, as previously discussed in Section \ref{sec: BSmodel}.

Following this, Dupire (1994) proposed a local volatility model, positing that the stock price follows a stochastic differential equation of the form: 
\begin{equation}
    dS_t = rS_tdt + \sigma(S_t,t)S_tdW_t
\end{equation}

where $r$ is the risk-free interest rate, $\sigma(S_t,t)$ is the local volatility function, and $W_t$ is a Brownian motion. 

In this thesis, we also denote the local volatility as $a$. If we set the risk-free interest rate $r = 0$, we obtain: 

\begin{equation}
    dS_t = a(S_t,t)S_tdW_t.
\end{equation}

Differing from the Bachelier model where $dS_t$ depends on $a(S_t,t)$ (the volatility of stock price at time t), the logistic model's $S_t$ depends on $a(S_t,t)S_t$ (the local volatility times the underlying security's price at time $t$).

In ``Additive logistic processes in option pricing"\cite{additivelog}, Carr and Torricelli, prove that risk-neutral distributions can be produced by simple no-arbitrage valuation formula, supported by additive processes. This laid out the foundation for the latest study on the application of convex duality. 

\subsection{Assumptions and Propositions}

In ``Convex Duality in Continuous Option Pricing Models", Carr and Torricelli's work pivots on several key assumptions, the foremost being the existence of a risk-neutral probability measure $\mathbb{Q}$. This measure implies that all non-dividend paying security prices are local martingales, thereby nullifying the possibility of arbitrage. Another key presumption is that the underlying security price process, denoted by $S = {S_t, t \geq 0}$, following a stochastic process with continuous sample paths, governed by the stochastic differential equation (SDE):

\begin{equation} \label{eq: 2.1}
    dS_t = a(S_t - s_0, t)dZ_t^{\mathbb{Q}}, \quad  t>0, \quad S_0=s_0 \in \mathbb{R}
\end{equation}

 for some $\mathbb{Q}$-Brownian motion $Z_t^{\mathbb{Q}}$.

A positive volatility function $ a: R \times R_+ \rightarrow R_+$ satisfies the following assumptions: \label{assumption}
        \begin{enumerate}
          \item (a) $a(\cdot,t) \in C^1(\mathbb{R},\mathbb{R}_+)$, for all $t \geq 0$;
          \item (b) $a$ satisfies a certain set of sufficient conditions for a unique martingale solution to equation \ref{eq: 2.1} to exist
          \cite{oksen2003} \cite{hirsch2011}.
          \item (c) The put price adheres to $p(\cdot, \cdot; s_0) \in C^{3,1}(R \times R_+, R_+)$ for all $s_0$. \label{assumption c}
        \end{enumerate}

Notice, the ``local volatility'' $a$ here depends on the price change $S_t - s_0$ at time $t$ of the underlying security during the time interval $[0, t]$, instead of only the price $S_t$ at time $t$. 

Let $\mathcal{G}_{s,t}^{\mathbb{Q}}$ be the infinitesimal generator of $(t, S_t)$:

\begin{equation} \label{eq: 2.2}
    \mathcal{G}_{s,t}^{\mathbb{Q}} := \frac{1}{2}a^2(s-s_0,t)\frac{\partial^2}{\partial s^2} + \frac{\partial}{\partial t}.
\end{equation}

Let $p(k, T; t, s, s_0)$ be the put option value function at time $t$ and stock price $s$, for a strike price $K$ at maturity date $T > 0$. By the fundamental asset pricing theorem introduced in Theorem \ref{thm: Fundamental Theorems of Asset Pricing}, we have

\begin{equation} \label{eq: 2.3}
    p(k, T; t, s, s_0) = E_t^{\mathbb{Q}}[(k-S_T)^{+}|S_t=s],
\end{equation}

where $E_t^{\mathbb{Q}}[\cdot|S_t=s]$ represents the conditional expectation, given $S_t = s$.

Recall $s$ stands for $S_t$, the stock price at time $t$, and $s_0$ represents the initial stock price. As $p(k,T;t,s,s_0)$ depends on both $s$ and $s_0$, which is governed by the PDE in equation \ref{eq: 2.1}, we can apply the KBE for the $p$, based on equation \ref{eq: 2.2}:

\begin{equation} \label{eq: 2.4}
        \mathcal{G}_{s,t}^{\mathbb{Q}}p(k, T; s, s_0) = 0, \quad k   \in \mathbb{R}, \quad t >0,
\end{equation}

with a terminal value:

\begin{equation} \label{eq: 2.5}
    p(k, T; s, s_0) = (k - S_T)^{+}.
\end{equation}

After utilizing Ito's lemma and Dupire's results in addition to the generator \ref{eq: 2.2}, we obtain the IVP:

\begin{equation} \label{eq: 2.11}
    \mathcal{G}_{k,T}^{\mathbb{Q}}p(k, T; s_0) = a^2(k-s_0,T)\frac{\partial^2 p}{\partial k^2}(k, T; s_0), \quad k \in \mathbb{R}, \quad T > 0,
\end{equation}

\begin{equation} \label{eq: 2.12}
    p(k, 0; s_0) = (k - s_0)^{+}.
\end{equation}

Under assumption (c), $\mathcal{G}_{k,T}^{\mathbb{Q}}$ becomes the space-time generator of the diffusion process:

\begin{equation}\label{2.13}
    dK_t = a(K_T - s_0, T)dW_T^{\mathbb{Q}}, \quad T > 0,\quad K_0 = k_0 \in \mathbb{R},
\end{equation}

for some Brownian motion $dW^{\mathbb{Q}}$ supported by the market filtration.

An excess strike-on-spot process $X_T$ is defined by

\begin{equation} \label{eq: 2.14}
    X_T := K_T - s_0
\end{equation}

and thus 

\begin{equation} \label{eq: 2.15}
    dX_T = a(X_T, T)dW_T^{\mathbb{Q}}, \quad T \geq 0, \quad X_0 = k_0 - s_0.
\end{equation}

In addition, by solving equation \ref{eq: 2.11} to $K_T$ applying the put option value function $p$ from equation \ref{eq: 2.12}, the authors introduce the stochastic process  $P = \{P_T, T \geq 0 \}$:

\begin{equation} \label{eq: 2.16}
    P_T := p(K_T, T; s_0), \quad T \geq 0, \quad P_0 = X_0^{+},
\end{equation}
here $P$ represents the put option value evolution on the randomized strike state variable. 

The process $\Delta = \{ \Delta_T, T \geq 0 \}$ is defined by:

\begin{equation} \label{eq: 2.17}
    \Delta_T := \frac{\partial p}{\partial k}(K_T, T; s_0),\quad  T > 0, \quad \Delta_0 \in \{0, 1\}.
\end{equation}

This presents the sensitivity of the put option value to the variation of the strike process K.

\begin{definition}[\textbf{Stochastic Process K}] \label{definition 1}
    The stochastic process $K$ is called the adjoint of the process $S$. In accordance, $\mathbb{P}$ and $\Delta$ are the adjoints of the processes $\{p(k,T;t,S_t,s_0)\}_{t\geq0}$ and $\frac{\partial p}{\partial s}(k,T;t,S_{t},s_{0})$ for $t \geq 0$ and will be referred to as the adjoint put price and the adjoint delta, respectively.

    The \textbf{adjoint} refers to the option values/sensitivities with respect to the strike variables.
\end{definition}

The adjoint processes $\Delta$ and $\mathbb{P}$ has the following connection:

\begin{proposition} \label{proposition 1}
    The processes $\Delta$ and $\mathbb{P}$ satisfy
    \begin{align}
    d\Delta_T = 
    &a(X_{T},T)((\frac{\partial a}{\partial x}(X_T, T)\frac{\partial^2}{\partial k^{2}} (K_{T},T;s_{0}))+a(X_{T},T\frac{\partial^3 p}{\partial k^3}(K_{T},T;s_{0})))dT \\
    &+ \frac{\partial^{2}p}{\partial k^{2}}(K_{T},T;s_{0})a(X_{T},T) dW_T^{\mathbb{Q}}, 
    \end{align}
    and
    \begin{equation}
    dP_{T} = d\langle X,\Delta \rangle_{T} + \Delta_{T} a(X_{T},T)dW_T^{\mathbb{Q}}.
    \end{equation}
\end{proposition}

Consider the associated stochastic exponential $\Sigma_T$:

\begin{align}
\Sigma_T & := \epsilon( \int_0^T - b(X_t,t) dW_t^\mathbb{Q})_T, \\
         & := \exp \left(- \int_{0}^{T} b(X_t,t) dW_t^\mathbb{Q} - \frac{1}{2} \int_{0}^{T} b^2 (X_t,t) dt\right), \quad T \geq 0  \\
solving \\
    d\Sigma_t & = -b(X_T,T)\Sigma_T dW_T^\mathbb{Q}, \quad with \quad \Sigma_0 = 1. 
\end{align}

According to Øksendal \cite{oksen2003}, $\Sigma_T$ is a local martingale on $[0,T^*]$. If it is a true martingale, then it induces an equivalent measure change to a measure $Q^* \sim Q$ on $(\Omega, \mathcal{F}_{T^*}^*, \mathcal{F}_{t\in[0,T^*]}, \mathbb{Q})$ through the Radon-Nikodym derivative:

\begin{equation} \label{eq: 2.26}
    \frac{d\mathbb{Q}^*}{d\mathbb{Q}} |_{\mathcal{F}_T} = \Sigma_T, \quad  for \quad T \in[0,T^*].
\end{equation}

Here we achieve martingale dynamics for $\Delta$ under $\mathbb{Q}^*$.

\begin{proposition}  \label{proposition 2}
Assume that $\Sigma$ is a true martingale. Then the adjoint delta $\Delta$ is a $\mathbb{Q}^*$-martingale, and its dynamics are given by
\begin{equation}
d\Delta_T = \frac{\partial^2 p}{\partial k^2}(KT,T;s_0) a(X_T,T) dW_T^{\mathbb{Q}^*}
\end{equation}
for some $\mathbb{Q}^*$-Brownian motion $W_{\mathbb{Q}^*}$.
\end{proposition}

Compared to the Black-Scholes-Merton model which changes from the actual probability measure $\mathbb{P}$ to the risk-neutral measure $\tilde{\mathbb{P}}$ (section \ref{sec: Risk-Neutral Measure}), this paper changes from the risk-neutral measure  into the dual delta measure.

\begin{definition}[\textbf{Dual Delta Measure}] \label{definition 2}
The measure $\mathbb{Q}^* \sim \mathbb{Q}$ is referred to as the dual delta measure.
\end{definition}

As discussed later in the paper\cite{ConvexDuality}, the dual delta measure approach is proved to be superior to the risk-neutral measure approach in several dimensions for both simplicity and complexity. Because of the nature that this paper was published in December 2022, future readings and research may emphasize evaluating both models further. 

To proceed, the authors propose a derivation of the Dupire IVP for pricing options with convex dual.

\begin{proposition} \label{proposition 3}
    Let $p(k, \tau)$ be the value of a put option with strike price $k$ and maturity $\tau$ written on a security $S$ following the dynamics (2.1), and let $p^*$ : $(0, 1) \times \mathbb{R}_+ \rightarrow [0, 1]$ be its convex dual in its first variable. Then $p^*$ is a solution of the PDE:
\end{proposition} 
    \begin{equation} \label{eq: 3.2}
                \frac{a^2 \delta (\delta , \tau )}{2}\frac{\partial^2 p^*}{\partial \delta^2} + \frac{\partial p^*}{\partial \tau} (\delta, \tau) = 0, \quad \delta \in (0,1), \quad \tau \geq 0, 
    \end{equation}
    with        $ p^*(\delta, 0) = 0$.

Subsequent to these developments, the research paper \cite{ConvexDuality} introduces the revised version of the previous assumptions \ref{assumption}, labeled ($a\prime$), ($b\prime$), and ($c\prime$):
\begin{enumerate} \label{assumptions revised}
    \item ($a\prime$) $a_\delta(\cdot, t) \in C^1([0, 1], (0, 1))$, for all $t \geq 0$;
    \item ($b\prime$) $a_\delta$ complies with a certain set of sufficient conditions for a unique martingale solution to equation $d\Delta^*_t = a_\delta(\Delta^*_t, T) dW^\mathbb{Q}_t$ to exist;
    \item ($c\prime$) a solution of equation \ref{eq: 3.2} exists such that $p^\ast \in C^{3,1}([0, 1] \times R_+, (0, 1))$.
\end{enumerate}

These updated assumptions serve as analogous conditions to ensure the coherence and validity of the theoretical framework in the paper.

\begin{definition}[\textbf{$\partial^*$, $X^*$, and $P^*$}]
We call $\partial^*$ the dual delta process, $X^*$ the dual excess price process, and $P^*$ the dual put value.

The process $X^{*} = \{ X_T^*, T \geq 0\}$ under $\mathbb{Q}^*$ in terms of the process $\Delta^{*}$, is defined by: 

\begin{align}
    X_T^{*} = \frac{\partial p^{*}}{\partial \delta}(\Delta_T^{*},T), \quad X_0^{*} = 0, \quad T \geq 0.
\end{align}

The stochastic process $P^* = \{ P_T^*, T\geq 0\}$ is defined as: 

\begin{align}
    {P_T}^* := p^{*} (\Delta_T^{*}, T), \quad P_0^{*} = 0, \quad T \geq 0
\end{align}

\end{definition}

In addition, it is assumed that under zero carrying costs and a risk-neutral measure, a put option’s underlying security price is a martingale. However, the ``adjoint delta” is not a martingale in the risk-neutral measure. A martingale setup can be restored after operating a measure change to a “dual” measure.
    
As an initial point of examination, under the classical mathematical theory of no-arbitrage, a put option's underlying security price is proposed to be a martingale under zero carrying costs and a risk-neutral measure. To illustrate, in a local volatility model where the state variable is an underlying security price following a driftless risk-neutral diffusion, the overlying put value process is demonstrated to continuously maintain its martingale properties in valuation time.

Contrary to this, the ``adjoint delta" — the derivative of the put value with respect to strike — does not retain its martingale characteristic under the risk-neutral measure. Nonetheless, a familiar martingale setup can be reintroduced by implementing a measure change to a ``dual" measure. In such a context, the adjoint delta will exhibit martingale properties, even though the put price does not. To identify an overlying process that inherently possesses the martingale property and thus can be interpreted financially, it becomes necessary to resort to the theory of convex conjugates. The key finding is that the dual delta and the Legendre transform of the put option value, under the dual delta measure, constitute a pair of martingale processes, aligned in an underlying-to-overlying relationship.

It is also assumed for this research that a risk-neutral probability measure $\mathbb{Q}$ exists, ensuring all non-dividend paying security prices behave as local martingales, thereby eliminating arbitrage possibilities.

Martingale's behaviors vary under specific conditions and assumptions, significantly contributing to the broader understanding of option pricing dynamics.

\subsection{Novel Contributions}

In \cite{ConvexDuality}, Carr and Torricelli proposed a slightly revised version of Dupire's Logistic model. In their model, the 'local volatility' depends on both the future price, $S_t$, and the initial price, $s_0$, instead of only on $S_t$."

The authors proved, for the logistic model, the dual delta is given by:
\begin{equation}
    \Delta_{\text{dual}}(K) = \frac{1}{\sigma}\ln\left(\frac{S}{K}\right) - \frac{r}{\sigma} + \frac{\kappa}{2\sigma^2}\left(1 - e^{-\sigma T}\right).
\end{equation}

They also proposed that the dual delta of a European put option with strike $K$ and maturity $T$ is given by:

\begin{equation}
    \Delta^*(K,T) = -\frac{\partial C(K,T)}{\partial K},
\end{equation}
    where $C(K,T)$ is the price of a European call option with strike $K$ and maturity $T$\cite{additivelog}.

In the case of the Bachelier model, the dual delta takes the form:
\begin{equation}
    \Delta_{\text{dual}}(K) = \frac{1}{\sigma}(S - K) + \frac{\kappa}{2}(T - t).
\end{equation}

The convex conjugate or Legendre transform, of the dual delta, where $V_{\text{dual}}$ is the value function associated with the dual delta, is given by:
\begin{equation}
    C_{\text{dual}}(x) = xK - V_{\text{dual}}(x).
\end{equation}

The following stochastic differential equation (SDE) describes the price dynamics of the asset:

\begin{equation} \label{eq: the price dynamics of the asset}
    \begin{split}
    dS &= S_t \sigma_t \sqrt{2b(t)b'(t)H(\delta(z))} dW_Q, \quad \\
    z &= \frac{\ln(S/S_0)}{b(t)}.
    \end{split}
\end{equation}

Carr and Torricelli \cite{ConvexDuality} developed the Logistic model as an advancement to the Bachelier model, aiming to find a new way to price options and strike a balance between ``complexity" and ``simplicity". The transformation between the dual and the primal, realized through the Legendre transform, the Convex Conjugate, corresponds to a one-to-one relationship between the values of put and put conjugate's boundary value problems (BVPs). The concept of convex duality leads to continuous martingale supporting logistic prices. The convexity implies that its corresponding blank is a negative quantity. The concept of conditional expectation is related to the pricing of a financial asset, reminding us of the essential probabilistic nature of financial markets.\cite{ConvexDuality}

In ``Convex Duality in Continuous Option Pricing Models", the Logistic model is named according to the unit scale cumulative distribution of a standard logistic random variable. 

Let's denote \textbf{the underlying security price} as $s_0 \in \mathbb{R}$ and let $k \in \mathbb{R}$ be the \textbf{strikes for call and put options}.

If we specify X to be a Bernoulli random variable with \textbf{success probability} 

\begin{equation}
    \delta \in (0, 1):
    H(X) = -\sum_{i=1}^{n} p_i \ln p_i.
\end{equation}

\textbf{The entropy function} of the Bernoulli random variable with success probability, $0 < \delta < 1$, is given by:

\begin{equation}
    H(\delta) = -\delta \ln \delta - (1 - \delta) \ln(1 - \delta).
\end{equation}

The first moment under the risk-neutral measure for the logistic model is given by:

\begin{equation}
    \eta_\delta(\delta) = H(\delta) \delta (1 - \delta), \quad \delta \in (0, 1).
\end{equation}

Carr and Torricelli \cite{ConvexDuality} assume that, for the dual delta, \textbf{the volatility function} $\boldsymbol{\eta(\delta)}$ is 
\begin{equation}\label{eq: the volatility function}
    \eta_\delta(\delta) = \sqrt{H(\delta)\delta(1 - \delta)}.
\end{equation}

\textbf{The dimensional put value} $\boldsymbol{p(k, s_0, T)}$ is defined as
\begin{equation} \label{eq: the dimensional put value}
    p(x, \tau)=b(\tau)\ln\left(1 + \exp\left(\frac{x}{b(\tau)}\right)\right).
\end{equation}

It can also be rewritten and thus introduces \textbf{the Option Pricing Formula} $\boldsymbol{p(k, s_0, T)}$ for the logistic model:
\begin{equation} \label{eq: Option Pricing Formula}
    p(k, s_0, T) = b(T)\ln\left(1 + \exp\left(\frac{k-s_0}{b(T)}\right)\right).
\end{equation}

According to the proposition from page 13, 

\begin{proposition} \label{proposition 5}
The value $p(k, T ; s_0)$ of a put option written on $S$ with MSV volatility given by $(4.1)$ equals
\[
\frac{k - s_0}{b(T)} p(k, T ; s_0) = b(T) \pi_{0} .
\]
Where $\pi(z)$ is the unique solution to the Neumann boundary problem
\begin{align}
    \begin{split}
        \eta^2 (z) \pi''(z) + z\pi'(z) - \pi(z) &= 0, \quad z \in \mathbb{R} \\
        \lim_{z\rightarrow -\infty} \pi'(z) &= 0, \\
        \lim_{z\rightarrow \infty} \pi'(z) &= 1,
    \end{split}
\end{align}
which is given by
\[
\pi(z) = \frac{1}{c_\pi} \left[ \exp \left(-\int_{-\infty}^{z} \frac{y \, dy}{\eta^2 (y)} + z \int_{0}^{z} \frac{1}{\eta^2(u)} du \right) \right] ,
\]
where
\[
c_\pi = \int_{-\infty}^{\infty} \exp \left( - \int_{0}^{y} \frac{u \, du}{\eta^2(u)} \right) \frac{dy}{\eta^2(y)} .
\]
\end{proposition}

\textbf{The put value function} $\boldsymbol{p(x,\tau)}$ is an elementary function of $\boldsymbol{x}$ and $\boldsymbol{b(\tau)}$, which is different from its version in the Bachelier model (the Normal model).

\textbf{The binary put price} $\boldsymbol{\delta(z)}$ is determined by

\begin{equation} \label{eq: the binary put price function}
    \delta(z) = \frac{1}{1 + e^{-z}}.
\end{equation}

\textbf{The variance function of the underlying security} $\boldsymbol{\eta(z)}$ is derived as:

\begin{equation} \label{eq: the variance function of the underlying security}
\begin{split}
\eta(z)     &   = \frac{\sqrt{ H(\delta(z))\delta(z)(1 - \delta(z))}}{\delta(z)(1 - \delta(z))}  \\
            &   = \sqrt{\frac{H(\delta(z))\delta(z)(1 - \delta(z))}{\delta(z)(1 - \delta(z))}} \\
            &   = \sqrt{(1+e^z)\log(1+e^{-z})+(1+e^{-z})\log(1+e^z)},
\end{split}
\end{equation}
\newline

To demonstrate the efficacy of the novel approach, they propose several volatility functions for the dual delta. These functions simultaneously exhibit realistic behavior and provide explicit formulas for valuing put options. 

They lay out some future research potentials, for example, it would be valuable to explore the duality approach in models where the dynamics of the underlying security are constrained to the positive half-line, aligning with the natural requirements of a price process. In such cases, the exponentiation of a normal random variable results in a log-normal random variable. However, to preserve the martingale property and compensate for the convexity of the exponential function, an introduction of negative drift becomes necessary in the original real-valued dynamics.

In contrast, in the paper, ``Additive logistic processes in option pricing", Carr and Torricelli demonstrated that within a logistic framework, a similar operation can be accomplished by raising the returns of the logistic cumulative distribution function (CDF) to an appropriate power\cite{additivelog}. This transformation leads to a skew-logistic random variable, giving rise to a ``log-skew-logistic" price distribution that corresponds to the conjugate power Dagum (CPD) model.

For the logistic model, the dual delta is given by:
\begin{equation}
    \Delta_{\text{dual}}(K) = \frac{1}{\sigma}\ln\left(\frac{S}{K}\right) - \frac{r}{\sigma} + \frac{\kappa}{2\sigma^2}\left(1 - e^{-\sigma T}\right).
\end{equation}

In the case of the Bachelier model, the dual delta takes the form:
\begin{equation}
    \Delta_{\text{dual}}(K) = \frac{1}{\sigma}(S - K) + \frac{\kappa}{2}(T - t).
\end{equation}

The \textbf{Convex Duality}, or \textbf{Legendre Transform} of the dual delta, where $V_{\text{dual}}$ is the value function associated with the dual delta, is given by:
\begin{equation}
    C_{\text{dual}}(x) = xK - V_{\text{dual}}(x).
\end{equation}

If we denote 

\begin{equation} \label{eq: pi delta}
    \pi^*(\delta) = -H(\delta),
\end{equation}

as the authors suggested, which produces the log-sum-exponential of 0 and z,

\begin{equation} \label{eq: log-sum-exponential of 0 and z}
    \pi(z) = ln(1+e^{z}).
\end{equation}

we can therefore derive the dimensional put value $p(k, s_0, T)$ as equation \ref{eq: the dimensional put value}.

The following stochastic differential equation (SDE) describes the price dynamics of the asset:
\begin{equation} 
    dS_t = S_t \sqrt{2\frac{b(t)b'(t)}{1-b(t)}\frac{H(\delta(z))}{\delta(z)(1-\delta(z))}} dW_t^Q, 
    \quad z = \frac{\ln(S_t/S_0)}{b(t)}.
\end{equation}

Here, $H$ stands for the Shannon entropy, $\delta$ is the standard logistic cumulative distribution function (CDF), and $b$ represents an increasing differentiable function with $\lim_{{t \to 0}} b(t) = 0$ and $\lim_{{t \to \infty}} b(t) = 1$. The authors also suggest future researchers may analyze the CPD diffusive model under the convex duality theory.

\subsection{Section Summary}

In summary, Carr and Torricelli introduce an innovative approach for valuing put options through the utilization of convex duality in this paper\cite{ConvexDuality}. Their methodology entails the identification of a pair of stochastic processes, under an appropriate measure, which act as dual counterparts to the put value and the underlying security price. They also derive the corresponding dual initial value problem (IVP) for the convex conjugate or Legendre transform of the put price. Instead of specifying the volatility of the underlying security price, their pricing framework requires the specification of the volatility of the put's dual delta. When the dynamics of the asset and dual delta can be separated into distinct time and space components, our approach yields consistent option prices that can be expressed using semi-closed forms.

A list, which enumerates the key notations and definitions, and a table, which summarizes the primal and dual specifications of the Bachelier and logistic models, are presented below.

\begin{itemize}
\item   $a$              :   the local volatility
\item   $k$              :   the strike for call and put options
\item   $z$              :   a standard logistic random variable
\item   $s_t$            :   the underlying security price at time $t$
\item   $S_t - s_0$      :   the price change of the underlying security price at time $t$
\item   $b(t)$           :   $a\sqrt{t}$
\item   $b(T)$           :   $\sqrt{T}$ if $a = 1$ 
\item   $N$              :   the standard normal distribution
\item   $p(x,\tau)$      :   the put value function
\item   $p(k, s_0, T)$   :   the dimensional put value, the option pricing formula
\item   $bp(k, s_0, T)$  :   the price of a binary put option
\item   $\eta(z)$        :   the variance function of the underlying security
\item   $\delta $        :   the success probability
\item   $\delta(z)$      :   the binary put price function, the 
\item   $H(\delta)$      :   the entropy function

    
\end{itemize}

    \begin{table}[H]
        \centering
        \caption{Summary of the primal and dual specifications of the Bachelier and logistic models.}
        \label{table: summary}
        \begin{tabular}{lll}
        \toprule
         & Bachelier & Logistic \\
        \midrule
        
        $\eta(z)$               & $a$                           & $\sqrt{(1 + e^z)\log(1 + e^{-z}) + (1 + e^{-z})\log(1 + e^z)}$ \\
        
        $\eta_\delta(\delta)$   & $N'(N^{-1}(\delta))$                   & $\sqrt{H(\delta)\delta(1 - \delta)}$ \\
        
        $p(k, s_0, T)$          & $(k - s_0)N(\frac{k-s_0}{b(T)})+b(T)N'(\frac{k-s_0}{b(T)})$  & $b(T)\ln\left(1 + \exp\left(\frac{k-s_0}{b(T)}\right)\right)$ \\
        
        $bp(k, s_0, T)$         & $N(\frac{k-s_0}{b(T)})$         & $(1 + \exp\left(-\frac{k-s_0}{b(T)}\right))^{-1} $ \\
        
        \bottomrule
        \end{tabular}
    \end{table}

Recall we assume the zero risk-free interest rate, with a risk-neutral probability measure $Q$, and zero dividend rate, which leads to local martingales for all security prices, from the underlying security of the put options, with no arbitrage.

The logistic model, according to Carr and Torricelli, is deemed ``simpler" than the Bachelier model in all aspects except for the specification of the primal underlying price variance. This implies that the logistic model is more easily manageable in terms of both inputs and outputs, offering advantages in various other respects.

In the next section, some numerical simulations and examinations are presented to help research and evaluate Carr's work, primarily focusing on Convex Duality in Option Pricing Models.

\newpage

\section{Numerical Simulation and Machine Learning for Convex Duality in Continuous Option Pricing} \label{sec: NSML}

In the following segment, we utilize Monte Carlo Simulation as a method for comparing and assessing different models. Specifically, we scrutinize the suggested Black-Scholes-Merton model, Bachelier model, and Logistic model using synthetic data. Furthermore, we establish an array of machine learning models to enrich our examination and prediction capabilities pertaining to option pricing, thus enhancing the scope of this research.

The special property of the financial market has ``jumps" such as the open and close times and holiday breaks for underlying securities, thus it is ideal to apply the concept of Continuous integral with Jumps when modeling for the convex duality seniors. 

Monte Carlo simulations are conducted, with simulated data, to compare the features' complexities of the logistic model with the Bachelier model. An analysis of runtime is provided, along with an evaluation of the simplicity.

\subsection{Monte Carlo Simulation for Models' Evaluation}


In Table \ref{table: summary}, formulas for some features of both the Bachelier and Logistic models are provided, and the Black-Scholes-Merton model is also introduced in Section \ref{sec: sto cal}. To analyze certain dimensions such as the simplicity, the complexity, and, most importantly, how well these models picture the dynamics and properties of the underlying security prices of the money market, the Monte Carlo Method is applied for data simulation, underlying security pricing, and numerical analysis based on the output results and the visualization images.

\subsubsection{Model Comparison Examination}

This thesis replicates the comparison, simulated by Carr and Torricelli, comparing the Bachelier model and the Logistic model, under the condition that 

\begin{equation}
    b(T) = \sqrt{T}.
\end{equation}

    Recall $b(T)$ is defined as $a\sqrt{T}$, where $a$ represents the local volatility, and $\sqrt{T}$ is the square root of the time $T$. 

For comparing purposes, we first visualize based on more commonly defaulted parameters: 

\begin{itemize}
    \item $E =  100$        : Strike price
    \item $T = 5$           : Time to maturity
    \item $r = 0.05$        : Risk-free rate
    \item $\sigma = 0.2$    : Volatility
\end{itemize}

for the generating of the below images.

\begin{figure}[H]
    \centering
    \includegraphics[width=0.5\textwidth]{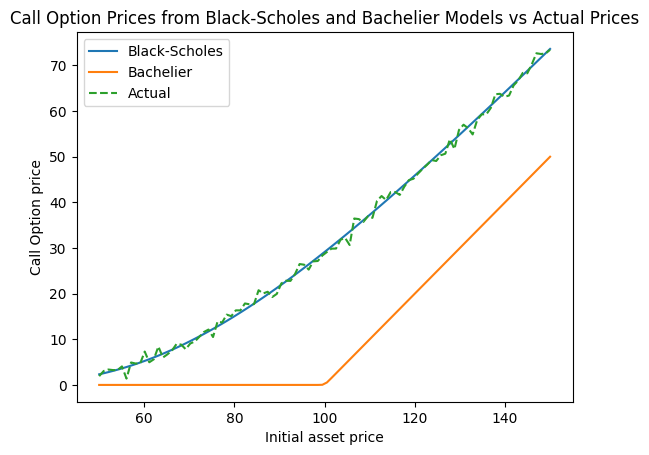}
    \caption{Call Option Prices from Black-Scholes and Bachelier Models vs Actual Prices}
    \label{fig:Call Option Prices from Black-Scholes and Bachelier Models vs Actual Prices}
\end{figure}

\begin{figure}[H]
    \centering
    \includegraphics[width=0.5\textwidth]{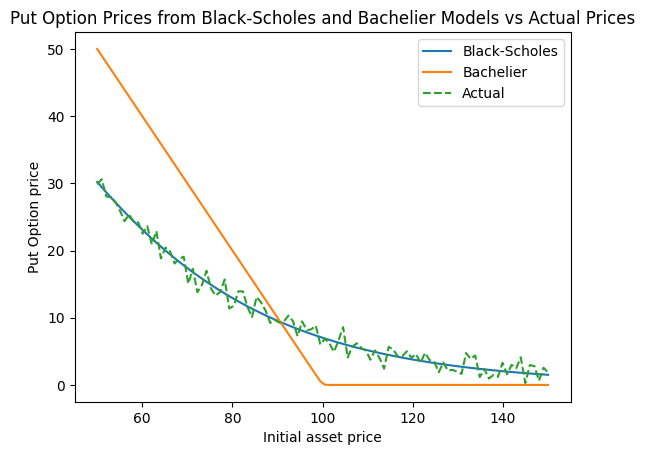}
    \caption{Put Option Prices from Black-Scholes and Bachelier Models vs Actual Prices}
    \label{fig:Put Option Prices from Black-Scholes and Bachelier Models vs Actual Prices}
\end{figure}

We can infer that the Bachelier model fits better for  the put option prices, especially when its put option price is close to the initial asset price. A more accurate replicate of Carr and Torricelli's proposal is examined in the following Section \ref{sec: observation and comments}.

To replicate their experiment, we need to set the local volatility $\sigma = a =1$ and $r = 0.00$, though this is a relatively unrealistic and nonideal scenario for such a high level of price fluctuation.

\begin{itemize}
    \item $E =  100$        : Strike price
    \item $T = 5$           : Time to maturity
    \item $r = 0.00$        : Risk-free rate
    \item $a = \sigma = 1$      : Volatility
\end{itemize}

\begin{figure}[H]
    \centering
    \includegraphics[width=0.5\textwidth]{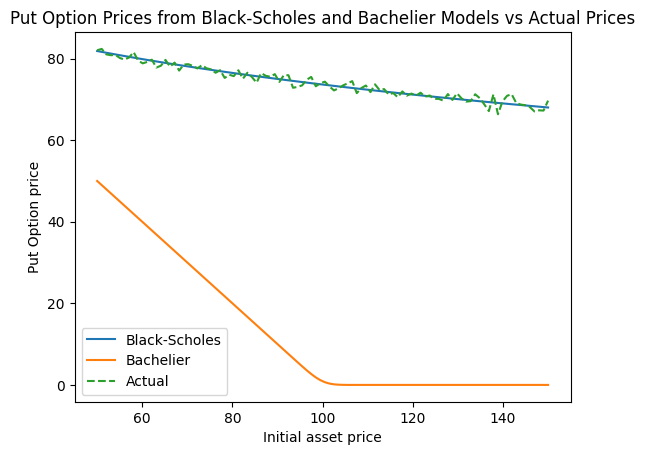}
    \caption{Put Option Prices from Black-Scholes and Bachelier Models vs Actual Prices}
    \label{fig:Put Option Prices from Black-Scholes and Bachelier Models vs Actual Prices 2}
\end{figure}

In this case, the predicted put option price based on the Bachelier model is far off compared to the Black-Scholes model and the actual prices. 

To compare the Bachelier and the Logistic model, we follow Carr and Torricelli's steps seeking to replicate the experiment.

\newpage

In their paper, ``Convex Duality in Continuous Option Pricing Models", Carr and Torricelli provide a visualization of the simulated results to illustrate how different features perform between these two models:

\begin{figure}[H]
    \centering
    \includegraphics[width=0.95\textwidth]{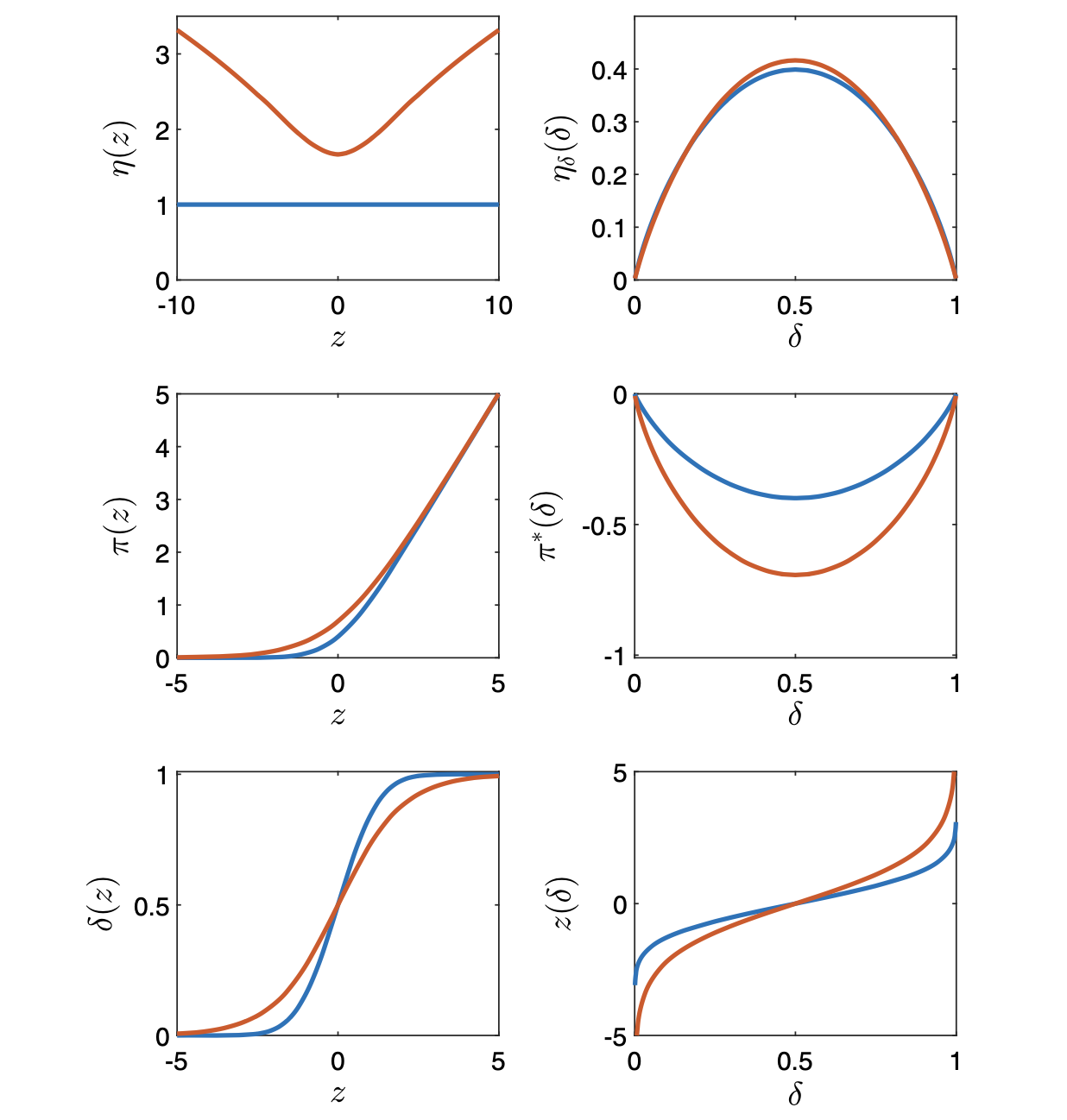}
    \caption{Comparisons of Functions in Bachelier and Logistic Models}
    \label{fig: Comparisons of Functions in Bachelier and Logistic Models}
\end{figure}

We replicate the simulation utilizing the equations and definitions provided in the previous section. However, some simulated results differ from Carr and Torricelli's work.

\begin{figure}[H]
    \centering
    \includegraphics[width=0.95\textwidth]{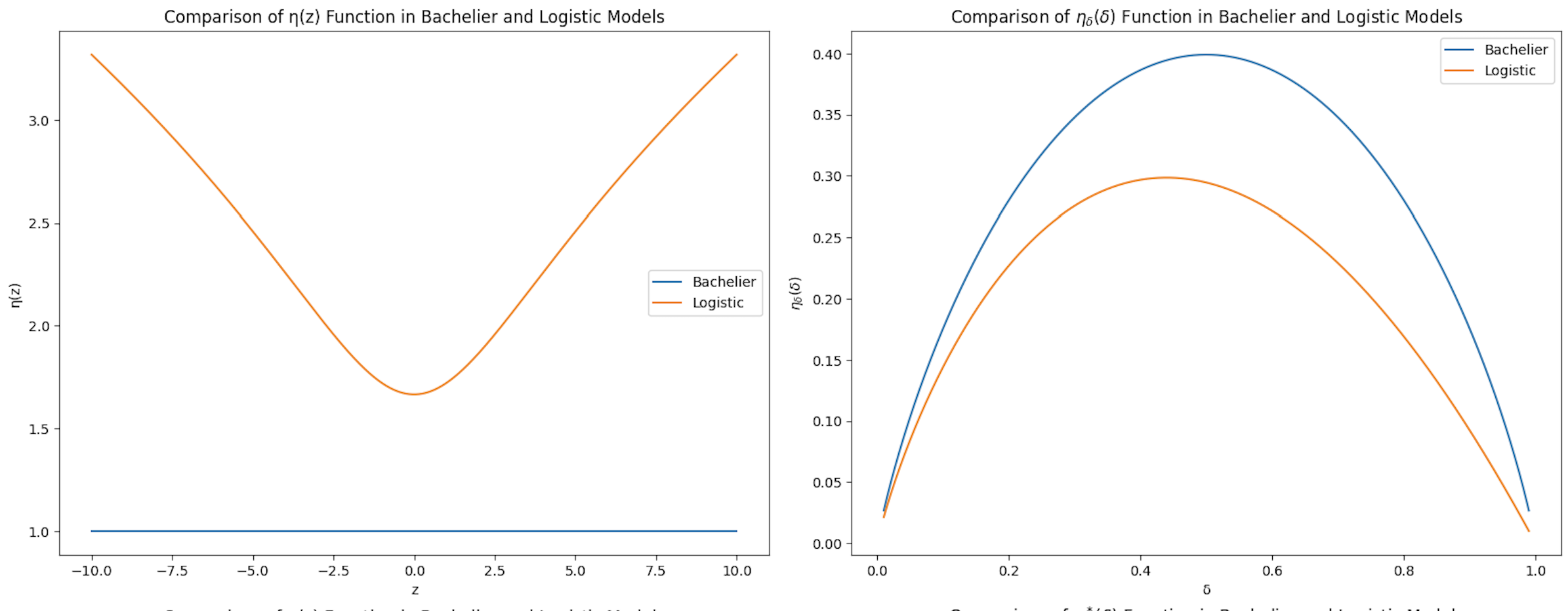}
    \caption{Examination of Comparisons in Bachelier and Logistic Models}
    \label{fig: Examination of Comparisons in Bachelier and Logistic Models}
\end{figure}

For example, for the $\eta_\delta(\delta)$ function comparison in figure\ref{fig: Examination of Comparisons in Bachelier and Logistic Models}, the difference in $\eta_\delta(\delta)$ values seems to disagree with the uniform texture presented in the original report. 

It is possible that there exist limitations to our numerical simulations, that are not perfect replicates of the original work. One of our future tasks is to find an explanation of the difference either by fixing our programming errors or providing a counterargument to their simulation.

\vspace{10mm}

We further discuss simplicity and complexity in the context of the normal model and the logistic model as introduced in Section \ref{sec: ConvexDuality}.

\subsubsection{Simplicity Comparison} \label{simplicity comparison}

Carr and Torrecelli state that the martingale density associated with the equivalent measure change from $Q$ to $\mathbb{Q}^*$, as well as its inverse, can be naturally expressed as the exponential of the stochastic integral of the initial spatial derivative of the dual delta variance rate. They conclude that this representation is more straightforward, thus simpler, than that in the strike coordinate given by the Bachelier model. It is infeasible for the Monte Carlo Simulation to model and evaluate the simplicity comparison argument due to its numerical experimenting nature. However, people can find their work and reasoning to be rigorous and convincing.

For the manageable context indicated in Table \ref{table: summary}, we agree with the authors' statements that the sole aspect in which the Bachelier model is considered "simpler" is the primal underlying price variance specification. However, in all other aspects, both regarding inputs and outputs, the logistic model proves to be more easily handled.

\subsubsection{Complexity Comparison} \label{complexity comparison}

The term \textbf{complexity} can be understood differently based on the context. In physical systems, it is defined as the measure of a system's state vector probability, distinct from entropy. In dynamical systems, statistical complexity refers to the size of the smallest program capable of statistically recreating the data set's patterns, offering a statistical description rather than a deterministic one. In mathematical studies of finite semigroups and automata, the concept of Krohn–Rhodes complexity is utilized. 

In this thesis, the complexity is examined numerically through the running time required for both models to execute the program.
    
Rounded to 5 digits, the running time comparison is presented in Table \ref{Bachelier vs. Logistic for Algorithm Complexity}. Notice, the simulated output values are just examples to illustrate the comparison method. 

\begin{table}[h!]
    \centering
    \caption{Bachelier vs. Logistic for Algorithm Complexity}
    \label{Bachelier vs. Logistic for Algorithm Complexity}
    \begin{tabular}{lccc}
    \hline
    \textbf{Function} & \textbf{Bachelier Cost t (s)} & \textbf{Logistic Cost t (s)} & \textbf{Better Model} \\
    \hline
        $\eta(z)$ & 0.00166 & 0.00356 & Bachelier \\
        $\eta_\delta(\delta)$ & 0.00212 & 0.00122 & Logistic \\
        $\pi(z)$ & 0.00125 & 0.00122 & Logistic \\
        $\pi^*(\delta)$ & 0.00193 & 0.00310 & Bachelier \\
        $\delta(z)$ & 0.00143 & 0.00132 & Logistic \\
        $z(\delta)$ & 0.00148 & 0.00149 & Bachelier \\
    \hline
    \textbf{Total} & 0.00987 & 0.01316 & Bachelier \\
    \hline
    \end{tabular}
\end{table}

\subsection{Machine Learning}


A financial application of Stochastic Calculus is trading underlying securities, such as options. It is a common trend for investment banks to develop algorithms utilizing machine-learning models to help forecast stock and option prices.  

Due to the theme of this thesis focusing on option pricing, not forecasting, we only introduce some previous works and research potentials of applying artificial intelligence to assist with financial mathematics problems.

Some sample machine learning modeling processes can be found in our previous papers ``Application of Convolutional Neural Networks with Quasi-Reversibility Method Results for Option Forecasting"\cite{CNN} and ``Optimizing Stock Option Forecasting with the Assembly of Machine Learning Models and Improved Trading Strategies" \cite{cao2022optimizing}.

A sample result table is provided to illustrate how applying different machine learning models can help improve pricing and forecasting option values given Quasi-Reversibility Method.

    \begin{table}[H]
        \centering
        \caption{Percentages of options with profits/ losses for different methods}
        \label{Percentages of options with profits/ losses for different methods}
        \begin{tabular}{lcc}
        \toprule
        Method & Profitable Options & Options with Loss \\
        \midrule
        QRM  & 55.77\% & 44.23\% \\
        Binary Classification & 59.56\% & 40.44\% \\
        Regression NN & 60.32\% & 39.68\% \\
        CNN Approach & 57.14\%  & 42.86\% \\
        \bottomrule
        \end{tabular}
    \end{table}

\subsubsection{Convex Duality and Convolutional Neural Network}

Recall we introduce Carr and Torricelli's recent works in Section \ref{sec: ConvexDuality}, because of the convex duality nature of the logistic model, applying linear programming seems to be a valid approach. Under such an assumption, we can treat the features, such as $S(t)$, $S(t) - S(0)$, $T - t$, $\Delta(t)$, $a$, as a matrix $A$ and therefore apply Convolutional Neural Network (CNN) to experiment with kernels $x$ to improve pricing option values $P$ through the change of measure from the risk-neutral measure into the dual delta measure.

    In a CNN, the idea of a dual delta could be used in a financial context to help predict option prices, which are particularly sensitive to changes in the underlying asset's price. In a CNN architecture, we could input asset price changes as part of the feature set to train the CNN model. Then, the model could potentially learn how these small changes impact the option prices and adjust its weights and biases (parameters) accordingly.

    The dual delta represents the rate of change of the option price with respect to changes in the price of the underlying asset. It can be viewed as a sensitivity measure, which is a valuable input for any financial model aiming to predict future option prices.

    Convex duality is frequently used in optimization problems, and in a sense, it is embedded in the process of training a CNN. The training process of CNNs involves solving an optimization problem (minimizing the loss function) that can sometimes be approached using convex optimization methods, depending on the choice of the loss function.
    
   For simplicity, an example of a 2D weight matrix $x$ of a CNN layer is represented as follows:

    \begin{equation*}
        \text{Weight Matrix} = 
        \begin{bmatrix}
        w_{11} & w_{12} & w_{13} \\
        w_{21} & w_{22} & w_{23} \\
        w_{31} & w_{32} & w_{33} \\
        \end{bmatrix}
    \end{equation*}

    In a practical application, these weights would be learned through a process of optimization, often using techniques such as backpropagation and gradient descent, to minimize the difference between the CNN's predictions and the actual values. The dual delta could play a role in this process by representing the sensitivity of the option prices to changes in the underlying asset prices, providing valuable information that CNN can use to adjust its weights and make more accurate predictions.

\subsubsection{Logistic Regression and Logistic Model}

    Logistic Regression, a statistical model, utilizes a logistic function for modeling a binary dependent variable. Similarly, the logistic model for implied volatility employs the logistic function, albeit for a different purpose. It models implied volatility as a function of moneyness and time to maturity. This function ensures that the volatility is a bounded and smooth function of its inputs, which aligns with the typical characteristics expected in the financial field. Both logistic regression and the logistic model for implied volatility exploit the logistic function's capacity to model a quantity (probability or volatility) expected to change in a non-linear, bounded manner as a function of its inputs. This underlines the versatility of the logistic function in diverse applications.
    
    Therefore, approaching option pricing and forecasting through Logistic Regression for the Logistic Model might have the potential for future research

\subsection{Section Summary}

In summary, Monte Carlo Simulation is applied to help numerically evaluate the results of Carr and Torricelli's. Moreover, machine learning, with rigorous proof and examination,  is recommended for improving the simulation and evaluation of models' pricing precision. A summary of observations, comments, limitations, and future research potential is presented in the following section.

\newpage
\section{Observations and Comments} \label{sec: observation and comments}

In summary, the first parts of the thesis serve as a survey of the basics of stochastic calculus for finance. The novel contributions of Peter Carr and Lorenzo Torricelli on ``Convex Duality in Continuous Option Pricing Models" are reviewed. Additional numerical simulation and machine learning models are utilized to initialize the evaluation experiments. 

The thesis commences with Section \ref{sec: history}, as an introduction to an encapsulated history of Financial Mathematics, providing the background and underscoring the significance of this field of study in the realm of financial applications.

Proceeding to Section \ref{sec: sto cal}, Stochastic Calculus for Finance, we progressively delve from fundamental concepts of the Binomial Asset Pricing Model into more complex and intricate theoretical perspectives embedded in the Continuous-Time Models, based on Steven E. Shreve's textbooks. 

The discussion evolves further in Section \ref{sec: ConvexDuality}, where we succinctly present Carr and Torricelli's paper, ``Convex Duality in Continuous Option Pricing Models". This section incorporates an introduction and literature review of the topic and previous work, key assumptions and conditions, and a snapshot of the model development process. It also sheds light on their novel contributions to the Logistic model for option pricing with convex duality.

Assumed no arbitrage and given $b(T) = \sqrt{T}$, let's denote $\boldsymbol{\delta^{-1}(z)}$ as the inverse unit scale cumulative distribution of a standard logistic random variable $z$:

\begin{equation} \label{eq: inverse unit scale cumulative distribution}
   \delta^{-1}(z) = 1 + e^{-z}.
\end{equation}

Additionally if we incorporate $\pi(z) = ln(1+e^{z})$ (equation \ref{eq: log-sum-exponential of 0 and z}). This results in an update on equation \ref{eq: inverse unit scale cumulative distribution}:

\begin{equation}
    \begin{aligned}
        \pi(z) &= \ln(1+e^{z}) = \ln(\delta^{-1}(-z)), \\
        \pi(-z) &= \ln(1+e^{-z}) = \ln(\delta^{-1}(z)).
    \end{aligned}
\end{equation}

These modifications result in a more precise and informative version of Table \ref{table: summary}.

    \begin{table}[H]
        \centering
        \caption{Summary of the primal and dual specifications of the Bachelier and logistic models, modified, given $b(T) = \sqrt{T}$.}
        \label{table: summary modified}
        \begin{tabular}{lll}
        \toprule
         & Bachelier & Logistic \\
        \midrule
        
        $\eta(z)$               & $a$                           & $\sqrt{\delta^{-1}(-z)\pi(-z) + \delta^{-1}(z)\pi(z)}$ \\
        
        $\eta_\delta(\delta)$   & $N'(N^{-1}(\delta))$                   & $\sqrt{H(\delta)\delta(1 - \delta)}$ \\
        
        $p(k, s_0, T)$          & $(k - s_0)N(\frac{k-s_0}{\sqrt{T}})+\sqrt{T}N'(\frac{k-s_0}{\sqrt{T}})$  & $\sqrt{T}\ln\left(1 + \exp\left(\frac{k-s_0}{\sqrt{T}}\right)\right)$ \\
        
        $bp(k, s_0, T)$         & $N(\frac{k-s_0}{\sqrt{T}})$         & $(1 + \exp\left(-\frac{k-s_0}{\sqrt{T}}\right))^{-1} $ \\
        
        \bottomrule
        \end{tabular}
    \end{table}

Looking towards future avenues, there is ample scope for additional research. The CPD diffusive model, for instance, could be analyzed under the lens of convex duality theory. A possibility also lies in providing an algorithmic proof that examines the simplicity and complexity interplay of the Logistic and Bachelier models. Furthermore, studies could focus on evaluating the superiority of specific contracts or portfolios in the context of option pricing under various theoretical and machine learning models.

In the ultimate Section \ref{sec: NSML}, Monte Carlo Simulation is employed as a practical tool for comparison and evaluation. We analyze the proposed Black-Scholes-Merton model, Bachelier model, and Logistic model in the light of artificially randomly generated data. Moreover, several machine learning models are suggested to extend the analysis and forecasting of option pricing, broadening the horizons of this study.

The numerical results provide an alternative approach to examining Carr and Torricelli's claims. Based on the Monte Carlo simulation, we numerically evaluate the complexity (section \ref{complexity comparison}), in terms of program running time, for option pricing of both models. A brief analysis of the simplicity comparison is presented in section \ref{simplicity comparison}.

    \begin{table}[H]
        \centering
        \caption{Bachelier vs. Logistic for Simplicity and Complexity}
        \label{Bachelier vs. Logistic for Simplicity and Complexity}
        \begin{tabular}{lc}
        \toprule
         & Superior Model\footnotemark \\
        \midrule
        Simplicity          &    Logistic \\
        Complexity          &    Bachelier \\
        \bottomrule
        \end{tabular}
    \end{table}    

 \footnotetext{The conclusion about the Bachelier model's superior complexity  performance (less running time) is based solely on our specific test scenario with simulated data. Results may vary with different datasets or algorithm designs.}

From a computational efficiency perspective, the Bachelier model exhibits superior performance in terms of overall run time. Specifically, it requires roughly 74.97\% of the time that the Logistic model necessitates, demonstrating its relative advantage in terms of algorithmic running time in this context. 

Lastly, for future research, machine learning is suggested to be utilized to assist in the simulation and evaluation of models' pricing precision. A comprehensive and meticulous mathematical investigation with the aim to ascertain the definitive veracity of these results awaits future research.

\vspace{10 mm}

The study of stochastic calculus for option pricing, incorporating aspects such as the Bachelier model, the Black-Scholes-Merton model, the Logistic model, convex duality, and numerical simulation, is to infinity and beyond.

\newpage
\section{Acknowledgment}

I wish to convey my deep appreciation to my supervisor, Professor Zhen-Qing Chen. His mentorship has been instrumental in both my academic and personal growth.


\newpage

\begin{thebibliography}{99}

\bibitem{bachelier1900}
Bachelier, L. (1900).
Théorie de la Spéculation
\textit{Annales scientifiques de l'École Normale Supérieure}.

\bibitem{black1973}
Black, F., \& Scholes, M. (1973).
The pricing of options on corporate liabilities.
\textit{The Journal of Political Economy}, 81, 637--654.

\bibitem{bowden2011}
Bowden, R. J. (2011).
Directional entropy and tail uncertainty, with applications to financial hazard and investments.
\textit{Quantitative Finance}, 11, 437--446.

\bibitem{CNN} 
Cao, Z., Du, W., \&  Golubnichiy, K.V. (2023).
Application of Convolutional Neural Networks with Quasi-Reversibility Method Results for Option Forecasting.
\textsc{Journal of Lecture Notes in Networks and Systems, Computing Conference 2023}.

\bibitem{cao2022optimizing}
Cao, Z., Guo, R., Du, W., Gao, J., \& Golubnichiy, K.V. (2023).
Optimizing Stock Option Forecasting with the Assembly of Machine Learning Models and Improved Trading Strategies.
\textit{To appear}. \url{arXiv preprint arXiv:2211.15912}.

\bibitem{carr2005}
Carr, P., \& Madan, D. (2005).
A note on sufficient conditions for no arbitrage.
\textit{Finance Research Letters}, 2, 125--130.

\bibitem{additivelog}
Carr, P., \& Torricelli, L. (2021).
Additive logistic processes in option pricing.
\textit{Finance and Stochastics}, 25, 689--724.

\bibitem{longtermrisk}
Carr, P., \& Torricelli, L. (2021).
Long Term Risk - A Time Change Approach.
\textit{SSRN}. \url{http://dx.doi.org/10.2139/ssrn.3995428}.

\bibitem{optionalitybinary}
Carr, P., \& Costa, D. (2022).
Optionality as a Binary Operation.
\textit{SSRN}. \url{http://dx.doi.org/10.2139/ssrn.4018065}.

\bibitem{stochasticvolatility}
Carr, P., Geman, H., Madan, D. B., \& Yor, M. (2002).
Stochastic Volatility for Lévy Processes.
\textit{Mathematical Finance}, 12(4), 345--382.

\bibitem{timechangedlevy}
Carr, P., Geman, H., Madan, D. B., \& Yor, M. (2002).
Time-Changed Lévy Processes and Option Pricing.
\textit{Bernoulli}, 8(6), 799--845.

\bibitem{ConvexDuality}
Carr, P., \& Torricelli, L. (2022).
Convex Duality in Continuous Option Pricing Models.
\textsc{Annals of Operations Research}.


\bibitem{guideBachelier}
Choi, J., Kwak, M., Tee, C. W., \& Wang, Y. (2022).
A Black-Scholes User's Guide to the Bachelier Model.
\textsc{Journal of Futures Markets}. \url{https://arxiv.org/abs/2104.08686}.

\bibitem{davis2007}
Davis, M. H., \& Hobson, D. G. (2007).
The range of traded option prices.
\textit{Mathematical Finance}, 17(1), 1--14.

\bibitem{dupire1994}
Dupire, B. (1994).
Pricing With A Smile.
\textit{Risk}, 7, 18--20.

\bibitem{Elliott}
Elliott, R. J. (1976) 
Stochastic integrals for martingales of a jump process with partially accessible jump 
times. 
\textit{Z. Wahrscheinlichkeitstheorie verw Gebiete}, 36, 213--226.

\bibitem{hirsch2011}
Hirsch, F., Profeta, C., Roynette, B., \& Yor, M. (2011).
Peacocks and associated martingales, with explicit constructions.
\textit{Springer}.

\bibitem{hull}
Hull, J. C. (2018).
\textit{Options, Futures, and Other Derivatives}.
Pearson.

\bibitem{itkin2018}
Itkin, A. (2018).
A new nonlinear partial differential equation in finance and a method of its solution.
\textit{Journal of Computational Finance}, 21, 1--21.

\bibitem{madan2002}
Madan, D. B., \& Yor, M. (2002).
Making Markov marginals meet martingales: With explicit constructions.
\textit{Bernoulli}, 8, 509--536.

\bibitem{merton}
Merton, R. C. (1973).
Theory of rational option pricing.
\textit{The Bell Journal of Economics and Management Science}, 4(1), 141--183.

\bibitem{oksen2003}
Øksendal, B. (2003).
Stochastic Differential Equations: An Introduction with Applications.
\textit{Springer}.


\bibitem{rockafellar1997}
Rockafellar, R. T. (1997).
Convex Analysis.
\textit{Princeton University Press}.


\bibitem{SCFI}
Shreve, S. E. (2004).
Stochastic Calculus for Finance I: The Binomial Asset Pricing Model.
\textit{Springer}.


\bibitem{SCFII}
Shreve, S. E. (2004).
Stochastic Calculus for Finance II: Continuous-Time Models.
\textit{Springer}.


\end{thebibliography}
\end{document}